\documentclass[12pt,preprint]{elsarticle}
\usepackage{hyperref,natbib,amsmath,amssymb,graphicx,multirow,algorithmic,color,microtype,fullpage,lineno}
\pdfoutput=1
\usepackage[skip=8pt]{caption}
\usepackage[title]{appendix}

\usepackage{graphicx}
\usepackage{array}

\usepackage{subfigure}

\usepackage[utf8]{inputenc}
\usepackage[T1]{fontenc}

\usepackage{amssymb, amsmath, amsfonts}
\usepackage{outlines}

\usepackage{enumitem}
\setlist[enumerate,1]{label=\Roman*.}
\setlist[enumerate,2]{label=\roman*.}
\setlist[enumerate,3]{label=\alph*.}

\journal{Computer Physics Communications}
\usepackage{xcolor}
\usepackage{listings}

\colorlet{mygray}{black!30}
\colorlet{mygreen}{green!60!blue}
\colorlet{mymauve}{red!60!blue}

\definecolor{webgreen}{rgb}{0,.35,0}
\definecolor{webbrown}{rgb}{.6,0,0}
\definecolor{RoyalBlue}{rgb}{0,0,0.9}
\definecolor{mywhite}{rgb}{1.0,1.0,1.0}

\definecolor{purp}{rgb}{0.4,0.2,0.8}

\hypersetup{
   colorlinks=true, linktocpage=true, pdfstartpage=3, pdfstartview=FitV,
   breaklinks=true, pdfpagemode=UseNone, pageanchor=true, pdfpagemode=UseOutlines,
   plainpages=false, bookmarksnumbered, bookmarksopen=true, bookmarksopenlevel=1,
   hypertexnames=true, pdfhighlight=/O,
   urlcolor=webbrown, linkcolor=RoyalBlue, citecolor=webgreen,
   pdfauthor={Jiayin Lu, Emanuel Lazar, and Chris H. Rycroft},
   pdfsubject={Multithreaded Voro++},
   pdfkeywords={Voronoi tessellation, computational geometry, multithreaded programming},
   pdfcreator={pdfLaTeX},
   pdfproducer={LaTeX with hyperref}
}

\newcommand{\R}{\mathbb{R}}

\newcommand{\vpp}{\textsc{Voro++}}
\newcommand{\votop}{\textit{VoroTop}}
\newcommand{\navg}{N_\text{avg}}
\newcommand{\nopt}{N_\text{opt}}

\renewcommand{\vec}[1]{\mathbf{#1}}
\newcommand{\co}[1]{\texttt{#1}}

\begin{document}
\begin{frontmatter}
\title{An extension to \vpp{} for multithreaded computation of Voronoi cells}
\author[SEAS,UW]{Jiayin Lu}
\author[BIU]{Emanuel A. Lazar}
\author[UW,LBL,SEAS]{Chris H. Rycroft}
\address[SEAS]{John A. Paulson School of Engineering and Applied Sciences, Harvard University, Cambridge, MA 02138, United States}
\address[UW]{Department of Mathematics, University of Wisconsin--Madison, Madison, WI 53711, United States}
\address[BIU]{Department of Mathematics, Bar-Ilan University, Ramat Gan 5290002, Israel}
\address[LBL]{Mathematics Group, Lawrence Berkeley Laboratory,
Berkeley, CA 94720, United States}

\begin{abstract}
  \vpp{} is a software library written in C++ for computing the Voronoi
  tessellation, a technique in computational geometry that is widely used for
  analyzing systems of particles. \vpp{} was released in 2009 and is based on
  computing the Voronoi cell for each particle individually. Here, we take
  advantage of modern computer hardware, and extend the original serial version
  to allow for multithreaded computation of Voronoi cells via the OpenMP
  application programming interface. We test the performance of the code, and
  demonstrate that it can achieve parallel efficiencies greater than 95\% in
  many cases. The multithreaded extension follows standard OpenMP programming
  paradigms, allowing it to be incorporated into other programs. We provide an
  example of this using the \votop{} software library, performing a
  multithreaded Voronoi cell topology analysis of up to 102.4 million
  particles.
\end{abstract}
\begin{keyword}
  Voronoi tessellation \sep computational geometry \sep multi-threaded programming
\end{keyword}
\end{frontmatter}

\section*{Program summary}

\noindent \textit{Program title:} \textsc{Voro++} \\ \vspace{-0.3em}

\noindent \textit{Developer's repository link:} \url{https://github.com/chr1shr/voro} \\  \vspace{-0.3em}

\noindent \textit{Licensing provisions:} BSD 3-clause (with LBNL modification) \\  \vspace{-0.3em}

\noindent \textit{Programming language:} C++ \\ \vspace{-0.3em}

\noindent \textit{External routines/libraries:} OpenMP \\ \vspace{-0.3em}

\noindent \textit{Nature of problem:} Multithreaded computation of the Voronoi tessellation
in two and three dimensions  \\ \vspace{-0.3em}

\noindent \textit{Solution method:} The \textsc{Voro++} library is built around several
C++ classes that can be incorporated into other programs. The two largest
components are the \texttt{container...} classes that spatially sort input
particles into a grid-based data structure, allowing for efficient searches
of nearby particles, and the \texttt{voronoicell...} classes that represent a
single Voronoi cell as an arbitrary convex polygon or polyhedron. The Voronoi
cell for each particle is built by considering a sequence of plane cuts based
on neighboring particles, after which many different statistics
(\textit{e.g.}\@ volume, centroid, number of vertices) can be computed. Since
each Voronoi cell is calculated individually, the Voronoi cells can be
computed using multithreading via OpenMP.

\section{Introduction}
\label{sec:intro}
\begin{figure}
  \begin{center}
    \begin{tabular}{cc}
      \includegraphics[width=0.4\textwidth]{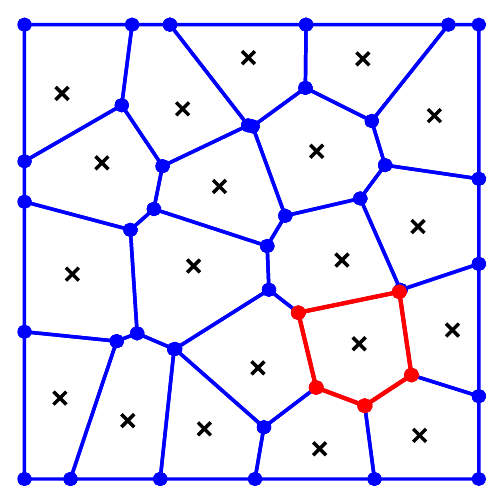} &
      \includegraphics[width=0.4\textwidth]{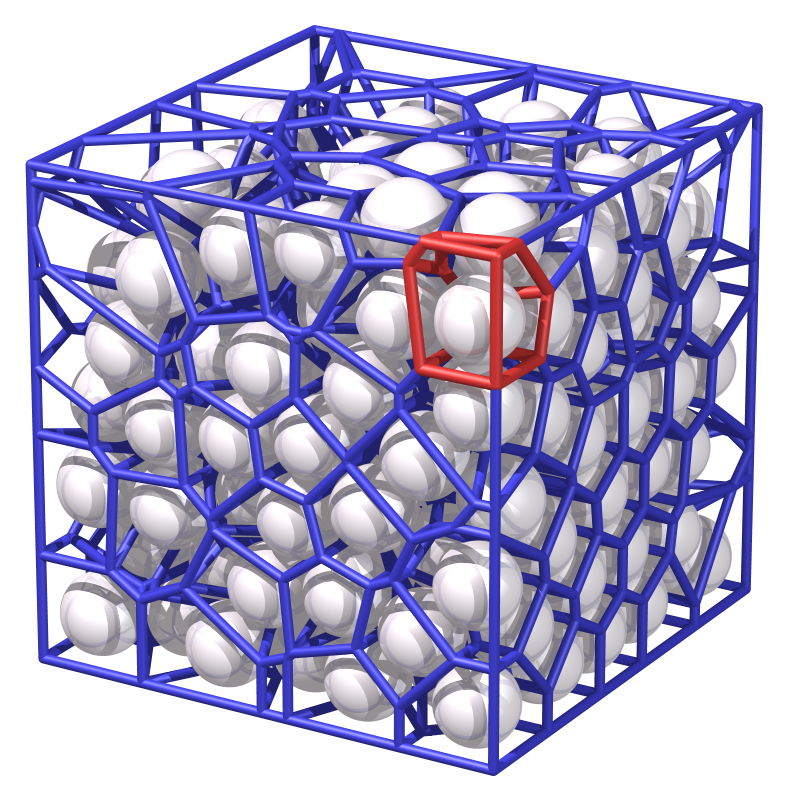} \\
      {\small(a)} & {\small (b)}
    \end{tabular}
  \end{center}\vspace{-0.5em}
  \caption{(a) An example two-dimensional Voronoi tessellation shown in blue
  generated by the black crosses. A single Voronoi cell in the tessellation is
  shown in red as an irregular polygon. (b) An example three-dimensional
  Voronoi tessellation shown in blue generated by the white spheres. A single
  Voronoi cell is shown in red as an irregular polyhedron.\label{fig:small}}
\end{figure}

The Voronoi tessellation was originally introduced in 1907~\cite{voronoi07} and
is now a broadly used technique in computational geometry~\cite{okabe09}.
Consider a set of points in a domain. Each point has a corresponding Voronoi
cell that is defined as the part of the domain that is closer to that point
than to any other. In two dimensions (2D) with the Euclidean metric, the Voronoi
cells are irregular polygons that perfectly partition the domain to create the
Voronoi tessellation (Fig.~\ref{fig:small}(a)). Each edge in the tessellation
is the perpendicular bisector between neighboring points. In three dimensions
(3D) the Voronoi cells are irregular polyhedra (Fig.~\ref{fig:small}(b)).
Voronoi cells can also be generalized to non-Euclidean geometries with
different distance metrics~\cite{chew85,klein88}.

The Voronoi tessellation has been used in a remarkable number of different
scientific fields. It has been extensively used to analyze systems of particles
or atoms, where features of the Voronoi cells (\textit{e.g.}\ volume, surface
area, number of faces) provide insight in particle structure; examples include
the analysis of granular materials~\cite{puckett11,guo14,rycroft12c},
colloids~\cite{rajaram12}, nanosphere systems~\cite{phillips10}, metallic
glasses~\cite{kramb13,gao14}, liquids~\cite{ruscher15}, as well as
active~\cite{wysocki14} and supercritical fluids~\cite{yoon2018topological,
yoon2019topological}. The Voronoi cells themselves, which form irregular
polygons/polyhedra, have been used to model different physical phenomena, such
as polycrystalline materials~\cite{benedetti13,orend15,gulizzi18},
solidification processes~\cite{budkewitsch94,feng19}, and biological
cells~\cite{murphy14}. The Voronoi tessellation has also been used to construct
computational meshes on which to solve partial differential
equations~\cite{guittet15,loubere10}, such as for climate
modeling~\cite{ringler08}, groundwater flow~\cite{freeman14,blanco-martin16},
and astrophysical flows~\cite{camps13}. Other applications include control of
multi-robot systems~\cite{teruel21}, calculating snow aggregate scattering
properties~\cite{honeyager16}, and modeling animal territorial
control~\cite{votel09}. There are many more examples than the ones given here,
highlighting the ubiquity of this geometrical construction~\cite{lazar2022voronoi}.

A variety of software packages are available for calculating the Voronoi
tessellation. The Qhull library~\cite{barber96,qhull_website} is widely used and
incorporated into MATLAB (via the \texttt{voronoin} command) and Python (via
the \texttt{scipy.spatial.Voronoi} command). The Computational Geometry
Algorithms Library (CGAL)~\cite{CGAL_website} provides a variety of functions
for computing the Voronoi tessellation, and
Triangle~\cite{shewchuk96,triangle_website} can compute the Voronoi
tessellation in 2D. All of these libraries primarily focus on computing the
Voronoi tessellation as an entire mesh, shown in blue in Fig.~\ref{fig:small}.

In 2009 Rycroft released \vpp{}~\cite{rycroft09c,voro_website}, a software library written
in C++ that takes the alternative approach of calculating the Voronoi cells
individually, so that each cell, an example of which is shown in red in
Fig.~\ref{fig:small}, is computed as a separate object. The library grew out of
research on particulate granular flows, where the Voronoi cell volumes were
useful for understanding particle packing
structure~\cite{rycroft06a,rycroft07,rycroft10}. The cell-based perspective has
some advantages and drawbacks compared to the entire mesh approach
(Subsec.~\ref{sub:cell_adv_dis}). However, it has proven effective in a wide
range in applications, particularly those involving rapid analysis of particle
systems. \vpp{} has a command-line utility that can perform a variety of
different analyses, and it has a C++ application programming interface (API)
that allows it to be called from user-written programs. It has been
incorporated into other software such as
LAMMPS~\cite{plimpton95,lammps_website} and
OVITO~\cite{stukowski10,ovito_website}.

\subsection{Algorithms for computing the Voronoi tessellation}
Since the 1970's, a wide variety of methods for computing the Voronoi
tessellation have been proposed~\cite{okabe09}. For computing
the entire Voronoi mesh, some popular methods include the Fortune sweeping
algorithm~\cite{fortune86,fortune87} and the incremental approach whereby the
mesh is continually updated as new particles are added~\cite{green78,lee80}.
Another method is to use the lift-up mapping, projecting a point $\vec{x}\in
\R^n$ to a paraboloidal surface $(\vec{x}, \|\vec{x}\|^2 ) \in \R^{n+1}$. The
hyperplanes tangential to the surface form facets that exactly match the
Voronoi tessellation when projected back to $\R^n$. This can be efficiently
computed in arbitrary dimensions using the quickhull algorithm~\cite{barber96},
which forms the basis of Qhull~\cite{qhull_website}. Another approach involves
introducing a computational grid, and sweeping out from each point with the
fast marching method~\cite{kimmel98,kimmel01} to construct Voronoi
cells~\cite{kimmel01a}, which is computationally expensive but more flexible
for calculations on spaces with non-Euclidean distance metrics.

\subsection{The cell-based approach: advantages and drawbacks}
\label{sub:cell_adv_dis}
The cell-based approach that \vpp{} uses has also been explored in the
literature~\cite{rhynsburger73,bentley80,boots83,quine84}. As
discussed by Okabe \textit{et al.}~\cite{okabe09} it has a significant
difficulty that is illustrated in Fig.~\ref{fig:cell_v_topo}(a,b). In most
cases, for randomly-distributed points, each vertex of the Voronoi tessellation
will be common between three Voronoi cells as shown in
Fig.~\ref{fig:cell_v_topo}(a). However, in certain situations a vertex may be
equidistant from four particles as shown in Fig.~\ref{fig:cell_v_topo}(b). This
could happen either because of a special arrangement of the particles
(\textit{e.g.}\ a crystalline formation), or for random arrangements when
particles happen to be aligned within the limit of floating point truncation
error. In the cell-based approach where the cells are computed independently,
small floating point errors in one cell may lead to the creation of additional
facets (shown by the red line in Fig.~\ref{fig:cell_v_topo}(b)) meaning that
the topologies of the edges and faces of the Voronoi cells are not consistent.

\setlength{\unitlength}{0.9bp}
\begin{figure}
  \vspace{-0.39in}
  \begin{center}
    \begin{picture}(490,120)
      \put(15,0){\includegraphics[scale=0.8]{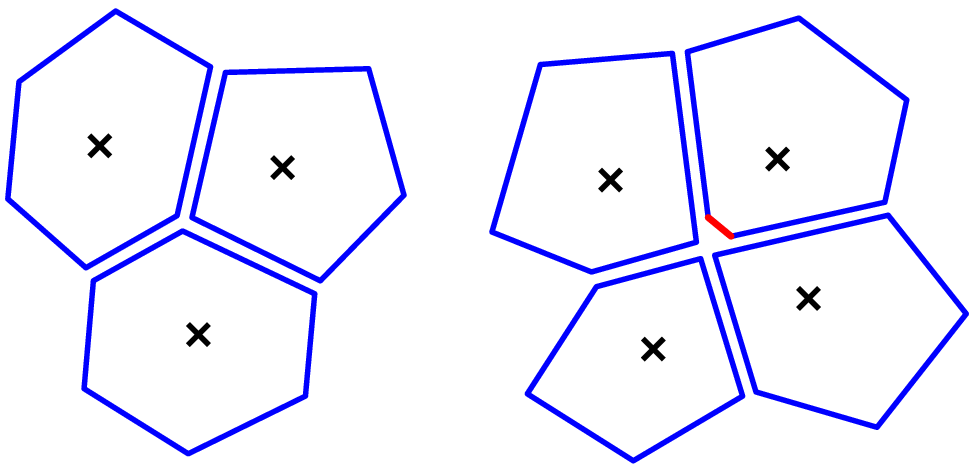}}
      \footnotesize
      \put(2,100){(a)}
      \put(140,100){(b)}
      \put(300,0){\includegraphics[width=2.3in]{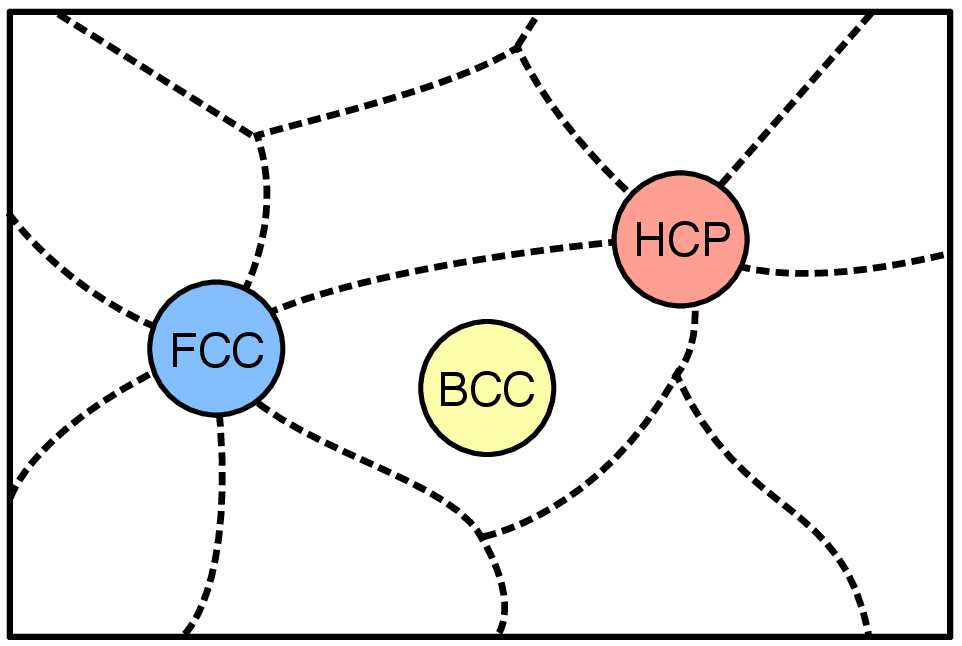}}
      \put(305,100){(c)}
    \end{picture}
  \end{center}
  \caption{(a) Typical case where three Voronoi cells (blue polygons) for three
  points (black crosses) meet at a vertex. The Voronoi cells would normally
  touch, but are spaced slightly apart from each other for illustrative
  purposes. (b) Special case where four Voronoi cells meet at a vertex that is
  equidistant from four points; floating-point errors could lead to additional
  small edges (red) for some cells. (c) Schematic representation adapted from
  Lazar \textit{et al.}~\cite{lazar15} where each dashed region represents a
  different Voronoi cell topology. Some common crystalline lattices
  (\textit{e.g.}\ BCC) are located within a single region, whereas others
  (\textit{e.g.}\ FCC, HCP) are located at junctions, and small perturbations
  from the ideal configuration sample different
  topologies.\label{fig:cell_v_topo}}
\end{figure}

Despite this difficulty, there are many situations where the cell-based
approach is attractive. The entire Voronoi mesh is often not required,
and the individual Voronoi cells can be analyzed independently. In \vpp{} a
typical workflow is to compute a Voronoi cell, calculate and store various
statistics about the cell, delete the cell, and move onto the next point.
Because only a single cell needs to be stored at any one time, this results in
a large memory saving and an improvement in cache efficiency, creating an
inherent performance boost over building the entire mesh. Furthermore, for
many commonly used measurements, it is not necessary for the edge topology to
agree perfectly. For example, Voronoi cell volumes and centroids (used in
Lloyd's algorithm~\cite{lloyd82,du99}) are not sensitive to small changes in
edge topology.

For other measurements, such as the number of faces or edges, the precise
topology of the Voronoi cell can have an appreciable effect. However, one can
reasonably argue that measurements that rely too heavily on these small
topological changes are problematic to begin with---it should not be the case
that a diagnostic indicator of a physical characteristic be sensitive to
truncation error (\textit{i.e.}~often around a factor of $10^{-16}$ in
double-precision arithmetic) particularly when experimental errors or
simulation discretization errors are usually far larger. The recent work of
Lazar, Srolovitz, and coworkers provides a useful theoretical framework in
which to address this issue~\cite{lazar15,lazar12,lazar13,leipold2016statistical,lazar2020voronoi}. Voronoi
cells can be pictured as residing in a phase space that is divided into
discrete regions representing the complete face and edge topologies
(Fig.~\ref{fig:cell_v_topo}(c)). Commonly studied crystalline lattices such as
FCC and HCP lie at intersections of these topologies, meaning that small
perturbations (\textit{e.g.}~thermal vibrations) around the idealized lattice
will push the Voronoi cells into a well-defined family of different topologies.
Other lattices such as BCC may lie in the interior of a single region. Lazar
released the \votop{} package~\cite{lazar17,vorotop_website}, which uses this
framework to analyze ensembles of Voronoi cell topologies, classify different
particle packings, and identify features such as grain boundaries. A key part
of \votop{} is the computation of the Weinberg vector~\cite{weinberg66} for
each Voronoi cell, which uniquely characterizes the cell's vertex and edge
topology.

\subsection{Outline of this paper}
\label{intro:paper outline}
Advances in supercomputing power have enabled simulations with very large
numbers of particles~\cite{phillips20,kondratyuk21}. In addition, there is
currently interest in developing data-driven approaches for screening large
databases of materials and structures, where the Voronoi tessellation can be a
useful analysis tool~\cite{willems12,pinheiro13,pinheiro13a,jalem18}. Thus
there is a need to compute the Voronoi tessellation at a large scale and in
parallel. Currently, there are some parallel approaches in the literature for
computing Voronoi cells in a distributed-memory
model~\cite{starinshak14,gonzalez16}. Here, we consider parallelizing \vpp{}
using a shared-memory model with multithreading. Since modern consumer laptops
and desktops contain CPUs with 4--8 cores and servers contain CPUs with upward
of 16 cores, multithreading enables a large practical speedup without the
additional complexity of using distributed-memory architectures.

The cell-based approach used by \vpp{} is inherently amenable to
parallelization, since each Voronoi cell can be computed independently. Here we
develop a general multithreaded extension of \vpp{} that provides good parallel
performance across a range of different scenarios. An ideal basis for doing
this is OpenMP, an API for shared-memory
multiprocessing~\cite{dagum98,openmp_website}. The core component of OpenMP is
a set of compiler directives beginning with \texttt{\#pragma omp} that instruct
the compiler to multithread certain lines and loops within a C++ code. A key
feature of OpenMP is that if a program is compiled without OpenMP enabled, then
the \texttt{\#pragma omp} directives are ignored and the compiler will create a
standard executable that runs in serial. For open-source scientific software,
which is compiled and run on a wide range of different systems, this serial
interoperability is a major advantage.

The extension to the \vpp{} API is designed to make it as simple as possible
for the user to incorporate multithreading into their programs. Multithreading
a loop over all Voronoi cells requires adding a small number of
\texttt{\#pragma omp} directives that match typical OpenMP usage. This has
required redesigning the mechanism for looping over Voronoi cells from previous
versions of \vpp{}, but this is done so that most of the complexity is hidden
from the user. Furthermore, we demonstrate that our extension is interoperable
with standard OpenMP functionality for tuning and controlling the division of
work between threads. With our extension, we show that we can achieve excellent
parallel efficiency of above 95\% across a range of cases.

\votop{} provides a particularly good example for our extension to \vpp{}. A
typical \votop{} analysis requires computing all of the Voronoi cells, and then
calculating the Weinberg vector for each one. The Weinberg vector is a
relatively expensive calculation, requiring $O(n^2)$ work for a Voronoi cell
with $n$ vertices. Thus, since the Voronoi cell and Weinberg vector
computations can be processed independently and divided evenly among threads,
this represents an ideal scenario for multithreading. In Section
\ref{sec:vorotop} we demonstrate this on a 36 core server. We show that the
time to process 102.4 million particles can be reduced from almost an hour to
under two minutes, highlighting a dramatic practical performance benefit.

\section{Methods}
\label{sec:methods}
\subsection{Overview of \vpp{}}
We now provide an overview of design and the key methods of
\vpp{}~\cite{rycroft09c}. The code is structured around several C++ classes for
storing particles and computing Voronoi cells. The code can perform both 2D and
3D computations, and the classes responsible for these have ``\co{\_2d}'' and
``\co{\_3d}'' suffixes, respectively. The algorithmic principles are
identical in both 2D and 3D. Therefore, for the rest of this section, we focus
on the 2D implementation for simplicity.

\subsubsection{The \co{container\_2d} class}
\label{sec:container_2d}
\begin{figure}
  \centering
  \includegraphics[width=0.3\textwidth]{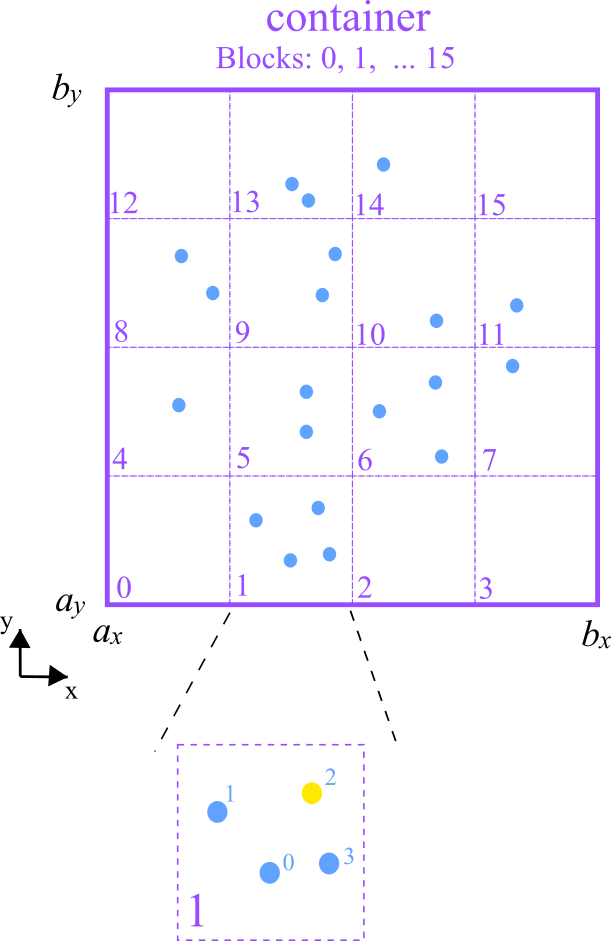}
  \caption{Illustration of a 2D container covering the rectangular region
  $[a_x,b_x]\times[a_y,b_y]$. The container is further divided into a $4\times
  4$ grid of blocks $(i,j)$ for $i,j \in \{0,1,2,3\}$. Each block $(i,j)$ is
  indexed as $k=i+4j$, so that $k\in\{0,1,\ldots,15\}$. Points are spatially
  sorted into blocks. The yellow point is the second point in block
  $1$.\label{fig:method_container}}
\end{figure}

\vpp{} has a variety of container types that represent rectangular domains
holding all of the positions. The \co{container\_2d} class holds particles in a
2D rectangle. The user can specify the coordinate ranges $[a_x,b_x]$ and
$[a_y,b_y]$, and indicate whether the container is periodic in each direction.
The container is further divided into a rectangular grid of $n_x\times n_y$
blocks of equal size, into which the particles are spatially sorted. In this
way, every particle can be identified by its block index, and its point index
inside the block. For example, in Fig.~\ref{fig:method_container}, the yellow
point is the $2^\text{nd}$ point in block $1$.

The grid of blocks provides a large boost in performance, allowing the code to
quickly locate neighboring particles during the Voronoi cell construction. See
Sec.~\ref{sec:blocks} for details on how the number of blocks is chosen. There
is also a variant container class called \co{container\_poly\_2d} that stores
polydisperse particle arrangements. Each particle has an associated radius,
which can be used to compute the radical Voronoi tessellation~\cite{okabe09}.

\subsubsection{Using a \co{voronoicell\_2d} class to compute a Voronoi cell}
\label{sec:using_cell}
The \co{voronoicell\_2d} class represents a single Voronoi cell as a convex
polygon, with a set of vertices connected by edges. The \co{voronoicell\_2d}
class contains routines for constructing the Voronoi cell, as well as routines
for computing different statistics about it, such as its area or centroid.

\begin{figure}
  \centering
  \includegraphics[width=1.0\textwidth]{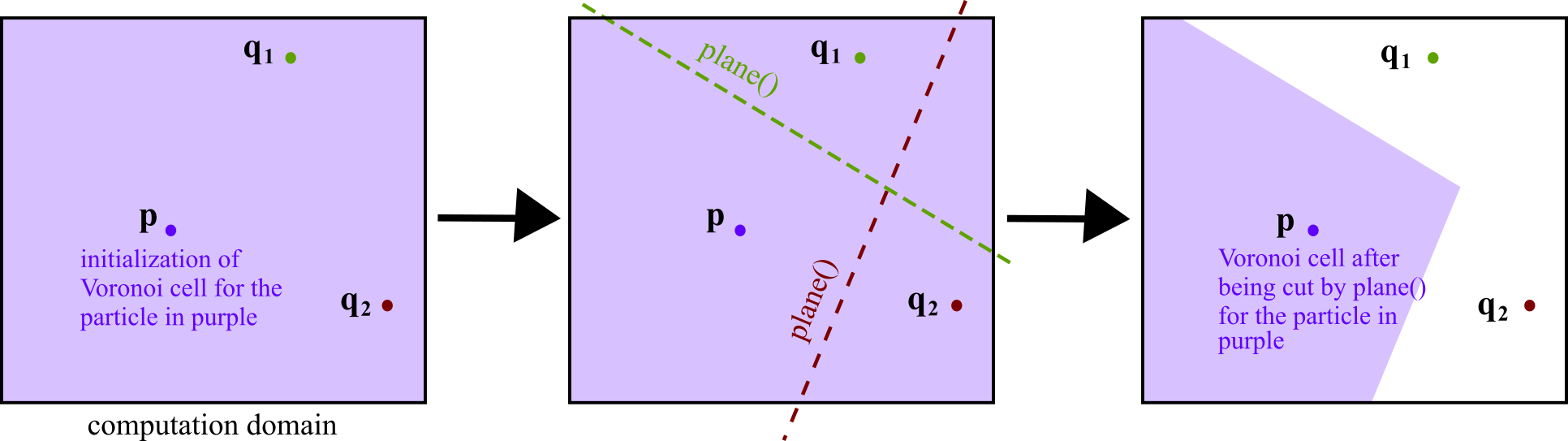}
  \caption{Illustration of the action of the \co{plane()} routine. The Voronoi
  cell of a particle $P$ located at $\vec{p}$ is initialized as a large rectangular box filling the
  computational domain. The \co{plane()} routine repeatedly cuts down the
  rectangular box by planes that are the perpendicular bisectors between the
  particle and its neighbors, located at $\vec{q_1}$ and $\vec{q_2}$.\label{fig:method_voronoicell}}
\end{figure}

The \co{voronoicell\_2d} class uses a coordinate system where the origin is
centered on the particle. Consider a specific particle $P$ located at position
$\vec{p}=(p_x,p_y)$ within the container. To compute its Voronoi cell, the
\co{voronoicell\_2d} class is first initialized as a rectangle
$[c_x,d_x]\times[c_y,d_y]$ filling the entire container, without considering
any other particles. Specifically, in the $x$ direction,
\begin{equation}
  [c_x,d_x] = \begin{cases}
    [a_x-p_x,b_x-p_x] & \qquad \text{if the $x$ direction is non-periodic,} \\
    [-\tfrac{b_x-a_x}{2},\tfrac{b_x-a_x}{2}] & \qquad \text{if the $x$ direction is periodic.}
  \end{cases}
\end{equation}
For the periodic case, the maximum extent of the initial Voronoi cell is
determined by the perpendicular bisectors of the periodic images of $P$ that
are displaced by $\pm(b_x-a_x,0)$. Similarly, in the $y$ direction,
\begin{equation}
  [c_y,d_y] = \begin{cases}
    [a_y-p_y,b_y-p_y] & \qquad \text{if the $y$ direction is non-periodic,} \\
    [-\tfrac{b_y-a_y}{2},\tfrac{b_y-a_y}{2}] & \qquad \text{if the $y$ direction is periodic.}
  \end{cases}
\end{equation}
To construct the Voronoi cell for $P$ the code then considers the effect of
neighboring particles. If a neighbor is located at $\vec{q}$ relative to $P$,
then that will remove the half-space
\begin{equation}
  \label{eq:hspace}
  \vec{r} \cdot \vec{q} > \frac{\vec{q}\cdot \vec{q}}{2}
\end{equation}
where $\vec{r}=(x,y)$. The boundary of the half-space, given by $\vec{r}
\cdot \vec{q} = \tfrac12 \vec{q} \cdot \vec{q}$, is the perpendicular bisector
between $P$ and its neighbor. The \co{voronoicell\_2d} class contains a
routine called \co{plane()} that recomputes the vertices and edges of the
Voronoi cell based on cutting by a plane. To compute the Voronoi cell for $P$,
the code considers the neighboring particles and applies plane cuts based on
removing the half spaces of the form given in Eq.~\eqref{eq:hspace}, as
illustrated in Fig.~\ref{fig:method_voronoicell}.

Hypothetically, if all plane cuts for all other particles are applied, then the
\co{voronoicell\_2d} class will precisely represent the Voronoi cell of $P$. In
practice, it is only necessary to consider plane cuts from a small set of
neighbors around $P$, as described in the following section.

\subsubsection{The \co{voro\_compute} class}
\label{sec:voro_compute}
The \co{container\_2d} class contains a member \co{vc} of class type
\co{voro\_compute\_2d}, which holds the data structures for computing the
Voronoi cells from the container's spatially sorted particles. For
a given particle $P$ the \co{voro\_compute\_2d} class computes the Voronoi cell
following the procedure in the previous section, but using as few plane cuts as
possible. As an example, consider the particle in the top left corner of
Fig.~\ref{fig:small}(a): its Voronoi cell has two faces that adjoin cells for
neighboring particles. Thus, if those two particles were considered first, then
all remaining plane cuts would have no effect.

\begin{figure}
  \centering
  \includegraphics[width=1\textwidth]{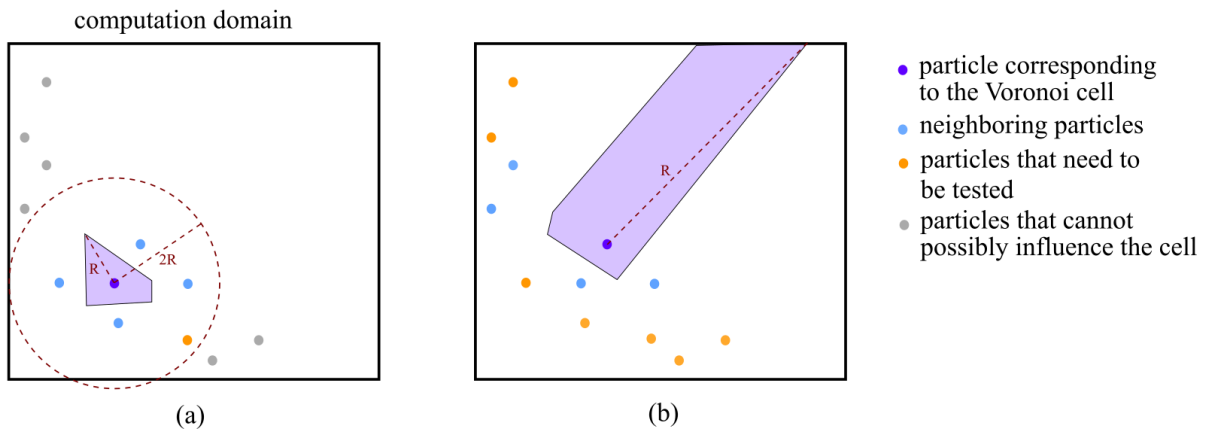}
  \caption{Illustration of different computational costs for different Voronoi
  cells. (a) A partially computed Voronoi cell is shown in light purple, after
  considering half-space intersections from four neighboring particles. The
  maximum distance to a vertex is $R$, and thus particles that lie outside a
  circle of radius $2R$ can be omitted from the computation; only a single
  additional particle needs to be tested. (b) If a particle does not have
  neighbors on all sides, its Voronoi cell may extend a long distance in one
  direction, so that $R$ is much larger. Hence many more particles lie within
  the circle of radius $2R$ and cannot be ruled out from the
  computation.\label{fig:method_vorocell_costs}}
\end{figure}

The \co{voro\_compute\_2d} class therefore computes Voronoi cells by
considering plane cuts from nearby points first, and then use bounds to
terminate the computation as soon as possible. The class first considers
particles in the same block as $P$, and then sweeps outwards to consider nearby
blocks. When each block is considered, two bounds can be used to determine
whether the Voronoi cell is complete or if more plane cuts are required:
\begin{itemize}
  \item Radius bound -- if $R$ is the maximum distance of a Voronoi cell vertex
    to $P$, then no particles more than a distance $2R$ away can possibly
    influence the cell. This bound is fast to compute, but it has no
    directional sensitivity: if a cell extends a long way in one direction then
    particles a long distance in other directions will still need to be tested.
  \item Block bound -- a given block in the grid can be tested to see if any
    particle within it can possibly influence the Voronoi cell. This can be
    done by performing a sequence of half-space intersection tests based on the
    block's corners. The code sweeps outward from $P$, testing blocks until it
    reaches those that cannot influence the cell. This computation is slower
    than the radius bound but it has directional sensitivity.
\end{itemize}
The \co{voro\_compute\_2d} class uses a combination of the two bounds. It
begins by using the radius bound, which works effectively for particles in
densely-packed regions with many close neighbors. This is illustrated in
Fig.~\ref{fig:method_vorocell_costs}(a), where the \co{voronoicell\_2d} polygon
is shown after considering four neighboring particles. At this point, the
bounding circle of radius $2R$ only contains a single additional particle.
Thus, once this particle is considered, then the Voronoi cell will be complete
and it will not be necessary to consider further particles.
In contrast, Fig.~\ref{fig:method_vorocell_costs}(b) shows a case where the
\co{voronoicell\_2d} polygon is extended in one direction because it is at the
edge of a particle arrangement. We refer to such cases as extended Voronoi
cells, where the maximum Voronoi vertex distance is large compared to a typical
inter-particle separation length. In this case, the circle from the radius
bound covers the whole domain, and no particles can be ruled out from the
computation.

If the radius bound is not successful in rapidly terminating the computation,
then the \co{voro\_compute\_2d} class switches over to the block bound. This
can help cut down the number of particles to consider, but the search space
that needs to be considered can still inherently be much larger. Hence extended
Voronoi cells can take substantially longer to compute. It is important to
consider this large difference in Voronoi cell computation time when designing
the multithreaded extension.

\subsubsection{Choice of the block size}
\label{sec:blocks}
The size of the grid of blocks in the container affects the computation
time. Let $N$ be the total number of particles and define $\navg = N/
(n_xn_yn_z)$ to be the average number of particles per block. If $\navg$ is too
large, then each block contains many particles, and since particles are not
spatially sorted within a block, the code must spend a long time looping
through all of them. If $\navg$ is too small, then the code must search through
many blocks to complete a Voronoi cell computation. The best performance
is achieved by choosing $\navg$ as a balanced value between these two extremes.

In addition, the optimal performance is usually achieved when the blocks have
roughly equal side lengths. Hence, for a given target average number of blocks,
$\nopt$, the code chooses the number of blocks as follows:
\begin{itemize}
  \item In 2D, set $\lambda = \sqrt{N/(\nopt(b_x-a_x)(b_y-a_y))}$ and define
    $n_x = \left\lceil \lambda (b_x-a_x) \right\rceil$ and $n_y = \left\lceil
    \lambda (b_y-a_y) \right\rceil$.
  \item In 3D, set $\lambda = \sqrt[3]{N/(\nopt(b_x-a_x)(b_y-a_y)(b_z-a_z))}$
    and define $n_x = \left\lceil \lambda (b_x-a_x) \right\rceil$, $n_y =
    \left\lceil \lambda (b_y-a_y) \right\rceil$, and $n_z = \left\lceil \lambda
    (b_z-a_z) \right\rceil$.
\end{itemize}
The use of the ceiling operator $\lceil \cdot \rceil$ ensures that the grid
dimensions are always greater than zero. For homogeneous random particle
arrangements, the best performance is achieved for $\nopt=3.4$ in 2D and
$\nopt=4.6$ in 3D, although the code is not that sensitive to this choice and
good performance is achieved across a wide range. These values are used by
default in the code, although they can be overridden by the user. In some of
the examples in this paper, which are chosen to highlight different scenarios,
we determine different values of $\nopt$ that improve performance.

\subsubsection{Procedure for Voronoi cell computation}
\label{procedure_cellCompute}
A typical usage of the library is as follows:
\begin{enumerate}
  \item Initialize the container. Insert particles into the container, spatially
    sorting them into the grid of blocks.
  \item Loop over the blocks in the container, and for each block:
    \begin{enumerate}
      \item Loop over the particles in the block, and for each particle:
        \begin{enumerate}
          \item Calculate the Voronoi cell of the particle.
          \item Compute and store required statistics about the Voronoi cell.
        \end{enumerate}
    \end{enumerate}
\end{enumerate}
Each Voronoi cell computation is independent of the others, making \vpp{}
highly suitable for parallel computation. A straightforward parallelization
approach is to distribute particles to different threads, and compute
their Voronoi cells simultaneously.

\subsection{Multi-threaded extensions}
\label{multithreaded_extensions}
\subsubsection{Changes to code architecture}
We now describe the changes required to make \vpp{} multithreaded using OpenMP.
The \co{vc} member within the container class, which is a \co{voro\_compute\_2d}
class, is responsible for calculating the Voronoi cell of a particle. The
\co{vc} member allocates workspace for searching through the blocks for
neighboring particles. Thus it is not thread-safe, since if two threads used
the same \co{vc} member, they would generate race conditions on the workspace.

Therefore we create copies of \co{voro\_compute\_2d} class object in the
container, based on the number of threads being used. \co{vc} is no longer a
single \co{voro\_compute\_2d} class, but becomes an array of them. The class
constructor accepts an additional argument \co{num\_t} that determines the
number of \co{voro\_compute\_2d} classes to allocate. Thread \co{k} then uses
\co{vc[k]} to compute its Voronoi cells. If needed, the function
\co{change\_number\_thread(num\_t)} can be used to reallocate the number of
\co{voro\_compute\_2d} classes available.

Moreover, in a typical loop like the one in Sec.~\ref{procedure_cellCompute},
the code creates a variable \co{c} that is a \co{voronoicell\_2d} class, for
representing the Voronoi cell of the particle. Similar to \co{vc}, different
threads in computation cannot use the same \co{voronoicell\_2d} object.
This issue is solved by creating thread-private copies of \co{c} in the parallel
computation.

\subsubsection{Random access iterator and OpenMP parallelization}
The OpenMP directive \co{\#pragma omp parallel} creates a team of threads to
execute a section of code. Within a parallel section, the directive
\co{\#pragma omp for} can be placed before a \co{for} loop, to distribute the
iterations of the loop to the different threads. Listing \ref{code:short_ex}
demonstrates how to use these two directives to parallelize the filling of an
array \co{c} with square roots of the integers.
\begin{lstlisting}[caption={Short example demonstrating basic OpenMP directives},label={code:short_ex}]
double c[256];
#pragma omp parallel
{
#pragma omp for
    for(int i=0;i<256;i++) {
        c[i]=sqrt(double(i));
    }
}
\end{lstlisting}
Each thread will be assigned a subset of values of \co{i} to set in the array.
Since each array entry can be set independently, this code can be multithreaded
without resulting in a race condition. In basic usage like the example above,
the \co{\#pragma omp for} directive is placed before loops over integers.
However, since version 3.0 of the OpenMP standard, it is possible to
parallelize a \co{for} loop using any C++ random access iterator.

\begin{figure}
  \centering
  \includegraphics[width=1\textwidth]{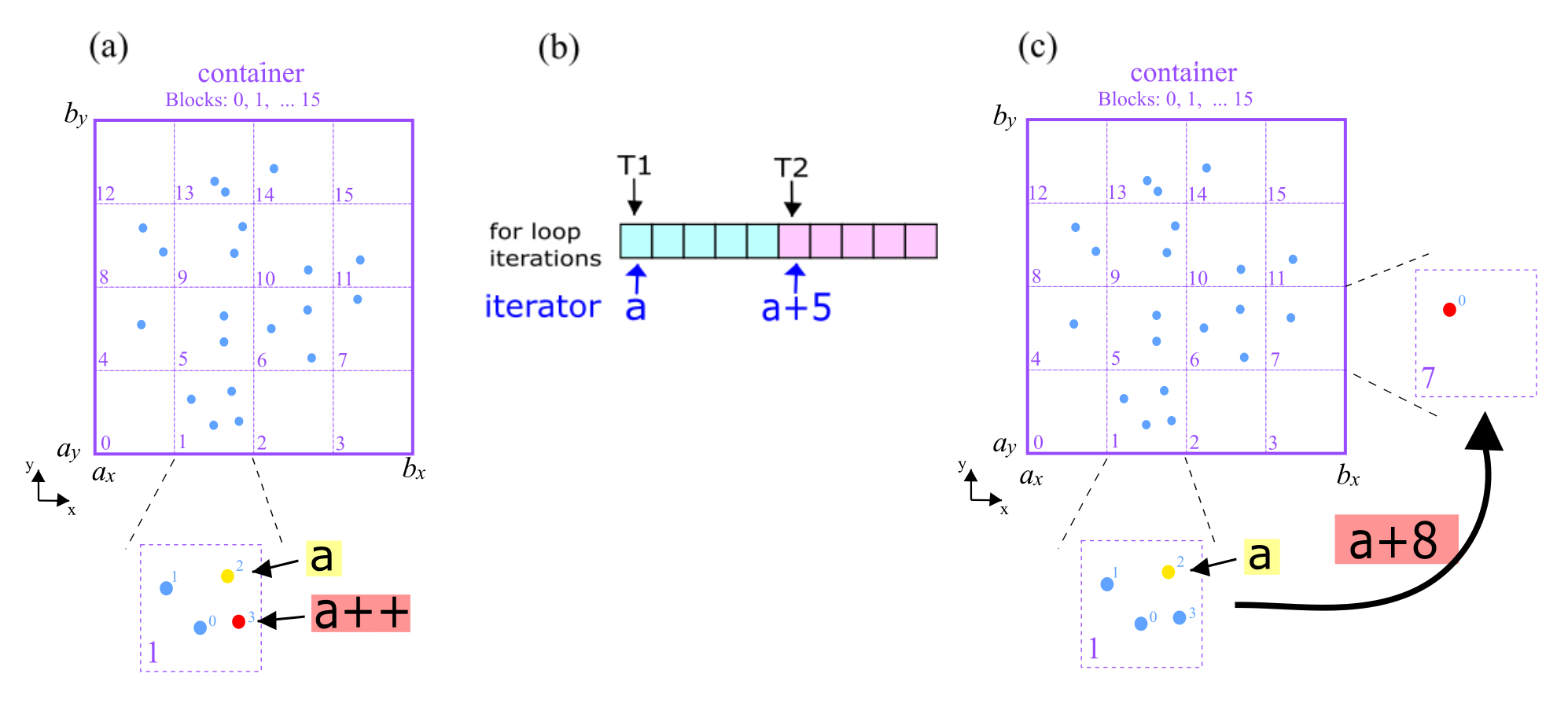}
  \caption{(a) Illustration of using a forward iterator to represent
  particles in the \co{container\_2d} class of \vpp{}. The forward iterator
  only allows the particle index to be stepped forward one at a time. If the
  iterator \co{a} is pointing at the yellow particle, then \co{a++} will step
  forward to the red particle. (b) Requirement of iterator \co{a} to represent
  particles in the container of \vpp{}. The iterator needs to be able to access
  particles at any arbitrary offset position relative to the current particle
  that the iterator is pointing at. For ten particles in the container and two
  parallel threads, thread $1$ starts at \co{a} and thread $2$ starts at \co{a+5}.
  (c) Illustration of using a random access iterator to represent particles in
  the container of \vpp{}. The random access iterator can access particles of
  any arbitrary offset positions relative to the current particle that it is
  pointing at. Here, \co{a+8} steps forward eight particles to the red
  particle.\label{fig:method_iter_overall}}
\end{figure}

In C++, each class can have associated iterator classes that are designed to
iterate over the elements of that class. There are several types of iterator,
differentiated by how much functionality they offer. The simplest is the
forward iterator, which supports basic operations for stepping forward
sequentially. The iterator \co{a} represents an index of the associated class,
and the forward iterator must support the operation \co{a++} to step forward to
the next index. Thus, in the context of a \vpp{} container, an iterator would
store the block index and point index of a particle, and the iterator would
support the operation \co{a++} to step forward to the next particle, as
illustrated in Fig.~\ref{fig:method_iter_overall}(a). Full requirements for the
functionality of a forward iterator are available in the C++
documentation~\cite{forward_iter_doc}.

In the multithreaded extension of \vpp{}, we created random access iterators on
all of the container classes to iterate over all of the particles. These
iterators support all of the functionality of a forward iterator, but also
contain additional functions for making arbitrary jumps in the particle
indexing. A key ability required by the random access iterator is to evaluate
\co{a+n} for an integer \co{n}, allowing the iterator to jump forward by \co{n}
steps in the index. This additional functionality is required since the threads
need to start at different points within a parallelized \co{for} loop
(Fig.~\ref{fig:method_iter_overall}(b)). Figure
\ref{fig:method_iter_overall}(c) illustrates how this works for a \vpp{}
container.

The \vpp{} iterators contain all the functions listed in the random access
iterator standard~\cite{randaccess_iter_doc}. This includes the dereference
operators \co{*a} and \co{a[n]}, which resolve to the member of the class the
iterator points to. However, in the current context, dereferencing is
conceptually unclear, since the Voronoi cell associated with the iterator does
not exist in memory and must be subsequently computed. Because of this, the
dereference functions within the C++ iterators simply emit errors if they
are called. For \vpp{}, the iterators are used to index into the container's
particles and loop through them, and thus the dereference operators are not
required in normal usage.

In addition to the standard iterator that loops through the particles, two
variations are provided: \co{iterator\_subset} that can loop over a subset of
particles, and \co{iterator\_order} that can loop over an ordered list of
particles.

\subsubsection{Example implementation}
An example of the multi-threaded version of \vpp{} is provided in Listing
~\ref{code:parallel_voro++}. The example demonstrates how to compute the
Voronoi cells of a random 2D particle arrangement, and then calculate their
average perimeter:
\begin{lstlisting}[caption={Example code of multi-threaded \vpp{}}, label={code:parallel_voro++},escapechar=|]
#include <cstdio>
#include <cstdlib>

#include "voro++.hh"
using namespace voro;

// Returns floating point number uniformly distributed over [0,1)
inline double rnd() {return (1./RAND_MAX)*static_cast<double>(rand());}

int main() {

    // Number of parallel threads
    int num_t=4;

    // Number of particles to use
    int N=100000;

    // Construct a 2D container as a periodic unit square divided into a
    // 160x160 grid of blocks. Each block initially holds up to 8 particles. The
    // final argument sets the number of voro_compute objects for use by the
    // threads.
    container_2d con(0.0,1.0,0.0,1.0,160,160,true,true,8,num_t); |\label{line:create container}|

    // Add particles to the container
    for(int i=0;i<N;i++) con.put(i,rnd(),rnd());    |\label{line:insert particles}|

    // Declare iterator                             |\label{line:use of iterator start}|
    container_2d::iterator cli;                     |\label{line:declare iterator}|

    // Parallel Voronoi computation to compute the average Voronoi cell
    // perimeter
    double tperim=0.;
#pragma omp parallel num_threads(num_t)
    {
        // Thread-private Voronoi cell object and perimeter counter
        voronoicell_2d c(con);                      |\label{line:create thread-private copy of voronoicell object}|
        double perim=0.;

        // Iterate through the particles
#pragma omp for                                     |\label{line:pragma omp directive}|
        for(cli=con.begin();cli<con.end();cli++)    |\label{line:loop through iterators}|
            if(con.compute_cell(c,cli))             |\label{line:compute voro cells in parallel region}|
                perim+=c.perimeter();

        // Add local perimeter counter to global perimeter counter using atomic
        // operation to prevent race condition
#pragma omp atomic
        tperim+=perim;                              |\label{line:atomic add}|
    }                                               |\label{line:use of iterator end}|

    // Print average Voronoi cell perimeter
    printf("Average Voronoi cell perimeter is %.12g\n",tperim/N);
}
\end{lstlisting}
After creating the container on line \ref{line:create container}, we then add
$N=10^5$ particles at random on line \ref{line:insert particles}. We then use
OpenMP on the random access iterator, and distribute particles to different
threads for parallel computation. Lines \ref{line:use of iterator
start}--\ref{line:use of iterator end} demonstrate using the iterator for
parallel computation. On line \ref{line:declare iterator} we declare an
iterator of the container. We then use the \co{for} loop on the iterator to
loop through all the particles in the container, as shown in lines
\ref{line:pragma omp directive} \& \ref{line:loop through iterators}. For each
particle, we compute its Voronoi cell (line \ref{line:compute voro cells in
parallel region}) and then add its perimeter to a local counter, \co{perim}.
Note that in certain situations (\textit{e.g.}\@ when occluded by a wall) a
Voronoi cell may not exist. In this case the \co{compute\_cell} function
returns false, and subsequent computation should be skipped.

For parallelization, we follow the standard OpenMP syntax. We mark the start of
the parallel block with \co{\#pragma omp parallel}, and add the \co{\#pragma
omp for} directive before the \co{for} loop. This simple implementation
following standard OpenMP and C++ practices is beneficial to the user.
At the end of the parallel block, on line \ref{line:atomic add}, each thread
adds its local perimeter counter to a global counter, \co{tperim}. Since the
threads are all writing to the global counter, this operation is performed
atomically to avoid a race condition. Once the parallel block has terminated,
the average perimeter is reported as $\approx 0.01265$. If the program is run
with different particle numbers $N$, one finds that the perimeter
asymptotically behaves like $4/\sqrt{N}$. Surprisingly, even though this is for
a random particle arrangement, this asymptotic value matches the perimeter of a
$\sqrt{N} \times \sqrt{N}$ grid of squares covering the domain.

\subsubsection{Load balancing}
\label{sec:load_balancing}
Load balancing is important for achieving good parallel performance. If the
iterations have even workloads, as shown in Table \ref{tbl:method_load}(a),
they should be assigned to threads evenly. In this way, all threads are busy
working during the full duration of the computation, and all threads finish at
approximately same time. In comparison, if the iterations are assigned unevenly
to threads, some threads will finish early and remain idle waiting for others
to finish, resulting in inefficient performance. On the contrary, if iterations
have uneven workloads, as shown in Table \ref{tbl:method_load}(b), then
assigning an equal number of iterations to each thread may result in
substantial idle time. Better performance can be achieved by distributing the
iterations so that the total workloads are balanced.

\begin{table}
  \begin{center}
\small
  \begin{tabular}{ |l||m{52mm}|m{52mm}|  }
 \hline
   & Assign workloads to threads evenly & Assign workloads to threads unevenly\\
 \hline
 (a) Even workloads
 & \begin{minipage}{0.18\textwidth}
  \vspace{3mm}
 \includegraphics[width=\linewidth]{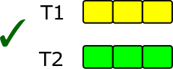}
 \end{minipage}
  \vspace{3mm}
 & \begin{minipage}{0.22\textwidth}
  \vspace{3mm}
 \includegraphics[width=\linewidth]{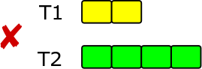}
 \end{minipage}
  \vspace{3mm}
 \\
 \hline
 (b) Uneven workloads
 & \begin{minipage}{0.3\textwidth}
  \vspace{3mm}
 \includegraphics[width=\linewidth]{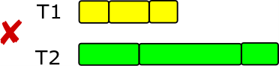}
 \end{minipage}
 \vspace{3mm}
 & \begin{minipage}{0.25\textwidth}
  \vspace{3mm}
 \includegraphics[width=\linewidth]{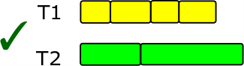}
 \end{minipage}
 \vspace{3mm}
 \\
 \hline
\end{tabular}
\end{center}\vspace{-0.8em}
\caption{Schematic illustration of load balancing among threads. (a) If
iterations have identical workloads, they should be assigned to threads evenly
for higher efficiency. (b) If iterations have uneven workloads, they should be
assigned to threads unevenly, following some strategies, so that the workloads
among threads are balanced.\label{tbl:method_load}}
\end{table}

\begin{table}
  \begin{center}
  \small
\begin{tabular}{ |m{3.5cm}||m{6.5cm}|c|  }
 \hline
 Parallel strategy & Traits & Schematic illustration\\
 \hline
 \co{schedule(static)} & 1. Iterations are pre-assigned evenly to threads. \newline 2. Good for identical workloads. \newline 3. Low overhead costs.
 & \begin{minipage}{0.25\textwidth}
 \includegraphics[width=\linewidth]{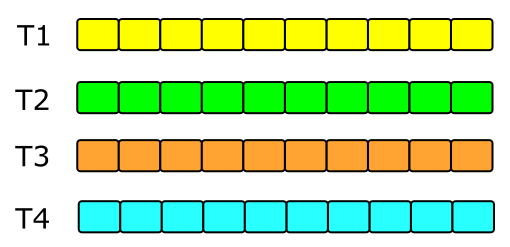}
 \end{minipage}
 \\
 \hline
 \co{schedule(dynamic)} & 1. Thread takes work as available. \newline 2. Good for imbalanced workloads. \newline 3. Higher overhead costs.
 & \begin{minipage}{0.3\textwidth}
 \includegraphics[width=\linewidth]{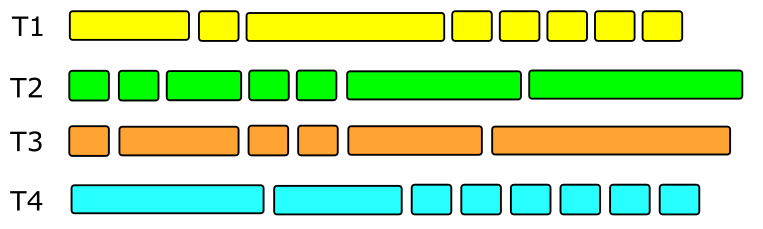}
 \end{minipage}
 \\
 \hline
 \co{schedule(guided)} & 1. ``Mixed'' strategy: Iterations are divided into chunks with decreasing chunk sizes, for threads to grab. \newline 2. Some imbalanced workloads. \newline 3. Higher overhead costs.
 & \begin{minipage}{0.3\textwidth}
 \includegraphics[width=\linewidth]{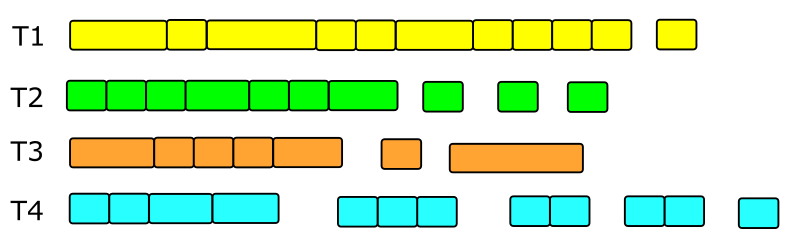}
 \end{minipage}
 \\
 \hline
\end{tabular}
\end{center}\vspace{-0.8em}
\caption{Three basic built-in OpenMP parallel strategies, their traits and
schematic illustrations of the work assignment
process.\label{tbl:OpenMP_built_in_strategies_main}}
\end{table}

OpenMP provides a number of strategies for load balancing. Here we explain
in detail the three most basic strategies, \co{schedule(static)},
\co{schedule(dynamic)}, \co{schedule(guided)}, with
Table~\ref{tbl:OpenMP_built_in_strategies_main} showing an overview of their
properties. For the \co{schedule(static)}, iterations (which for \vpp{} correspond to
particles) are pre-assigned to the threads by dividing the entire set of work
units into equal sections. This strategy is good for identical workloads, and
it has low overhead costs, since once the work is assigned, each thread can
operate independently. For the \co{schedule(dynamic)}, each thread takes a unit
of work as available and is never idle. If a thread finishes its work, it
immediately takes the next iteration available. This strategy comes with higher
overhead costs since the threads must coordinate with each other when taking
more work. \co{schedule(guided)} is a mixed strategy. Threads take work as
available and are never idle, but rather than taking one unit of work per time,
each thread takes a chunk. Moreover, the chunk sizes decrease as the assignment
progresses. This is good for when there is some imbalance in workloads, but
does not offer as much flexibility as \co{schedule(dynamic)}. It also has high
overhead costs.

Any of these strategies can be used with the multi-threaded \vpp{} extension,
by adding the corresponding keyword after the \co{\#pragma omp for} directive.
For example, in Listing~\ref{code:parallel_voro++}, changing
line~\ref{line:pragma omp directive} to \co{\#pragma omp for schedule(dynamic)}
will enable the dynamic assignment. In addition, there are two modified
strategies that accept an additional integer \co{chunk} option. In
\co{schedule(static,chunk)} the threads each take \co{chunk} work units at once
to process, rather than dividing up the total work into even sections. This
still does not require any thread coordination, but affects the work
distribution.\footnote{For example, consider two threads operating on the
integers $(0,1,\ldots,15)$. With \co{schedule(static)} the work is given to the
two threads as $(0,1,\ldots,7)$ and $(8,9,\ldots,15)$. With
\co{schedule(static,3)} the first thread receives the chunks $(0,1,2), (6,7,8),
(12,13,14)$ and the second thread receives the chunks $(3,4,5), (9,10,11),
(15)$ with the final chunk being truncated.} In
\co{schedule(dynamic,chunk)} the threads take \co{chunk} units of available
work at once. This cuts down on the overhead costs, since there are fewer
points where the threads must coordinate. Depending on the arrangement of
particles, one strategy may be preferred over another. In particular,
Sec.~\ref{sec:voro_compute} showed that for inhomogeneous particle
distributions, the Voronoi cell computation can vary by a large factor. In this
case, we expect that the dynamic or guided scheduling strategy will be
advantageous.

\subsubsection{Parallel insertion of particles}
\label{sec:parallel_put}
In Listing~\ref{code:parallel_voro++} the time spent on computing the Voronoi
cells is much larger than the time spent on inserting the particles into the
container. Nevertheless, once the particle system grows larger, the time for
particle insertion becomes substantial. Because of this, we also developed an
approach for inserting particles into the container in parallel.

As shown on line \ref{line:insert particles} of Listing
\ref{code:parallel_voro++}, the \co{put(...)} function is used to insert a
particle into the container in serial. Internally, this function first computes
the block index \co{k} that the particle lies within. The container class
has two array entries associated with the block: \co{co[k]} is the total number
of particles currently in the block, and \co{mem[k]} is the total number of
memory slots allocated for the block, typically initialized to 8 at the start
of the computation. The \co{put(...)} function therefore does the following:
\begin{enumerate}
  \item If \co{co[k]==mem[k]} then all memory slots are full. In this case,
    dynamically reallocate the memory for this block, and double the available
    slots.
  \item Add the particle information in the \co{co[k]} slot, and then increment
    \co{co[k]}.
\end{enumerate}
This routine is not thread-safe for two reasons: (a) if two threads try to
reallocate the memory simultaneously it will be corrupted, and (b) if two
threads try and increment \co{co[k]} simultaneously this will create a race
condition.

In the multithreaded extension, we created a function \co{put\_parallel(...)}
that can insert particles into the container in a thread-safe manner. This
function also requires an overflow buffer for storing some particles. The
\co{put\_parallel(...)} does the following:
\begin{enumerate}
  \item \label{ite:pp1} Use the \co{\#pragma omp atomic capture} directive on the operation
    \co{l=co[k]++}. This function will atomically increment \co{co[k]} while
    storing its previous value in the variable \co{l}.
  \item \label{ite:pp2} If \co{l>=mem[k]} there is no available memory slot.
    Store the particle in the overflow buffer.
  \item \label{ite:pp3} If \co{l<mem[k]}, store the particle into slot \co{l}.
\end{enumerate}
In step \ref{ite:pp1}, the atomic capture directive ensures that each thread
will obtain a unique slot number to write into. For example, if $\co{co[k]}=2$
and two threads insert a particle into this block, then one is guaranteed to
obtain $\co{l}=2$ and the other is guaranteed to obtain $\co{l}=3$. Afterward
$\co{co[k]}=4$.

Within \co{put\_parallel(...)}, it is not possible to dynamically reallocate
the memory, since this could interfere with other threads. Hence, if a block
has no available slots, it must be stored into the overflow buffer. Assuming
that the block memory was roughly allocated correctly beforehand, the number of
particles going to the overflow buffer should be a small fraction of the total.
The overflow buffer operations are done within an OpenMP \co{critical} section
of code, so that only one thread operates on the buffer at one time. If the
overflow buffer runs out of space, this thread can dynamically extend the
buffer without causing a race condition. Once all of the particles have been
added with \co{put\_parallel(...)}, it is necessary to call a function
\co{put\_reconcile\_overflow()}. This function operates serially on the
overflow buffer, and dynamically extends the block memory as needed to ensure
there are enough available slots to insert any remaining particles.

Using parallel insertion can cause the code to produce slightly different
results when it is repeated. Depending on the exact timing of the threads,
which differs from run to run, the particles may be inserted into the blocks in
a different order. This reordering can have a minor effect on the floating
point errors that are incurred during the Voronoi cell construction. The code
also contains a function \co{add\_parallel(pt\_list,num,nt\_)} that takes in a
pointer \co{pt\_list} to array of \co{num} particle positions stored as
$(x,y,z)$ triplets, with an additional radius argument for the polydisperse
containers. It performs a multi-threaded insertion of the particles into the
container using \co{nt\_} threads.

\section{Parallel Performance}
\label{sec3:parallel performance:main header}
\subsection{Voronoi cell computation}
\label{sec:parallel_performance}
\begin{figure}
  \centering
  \includegraphics[width=1\textwidth]{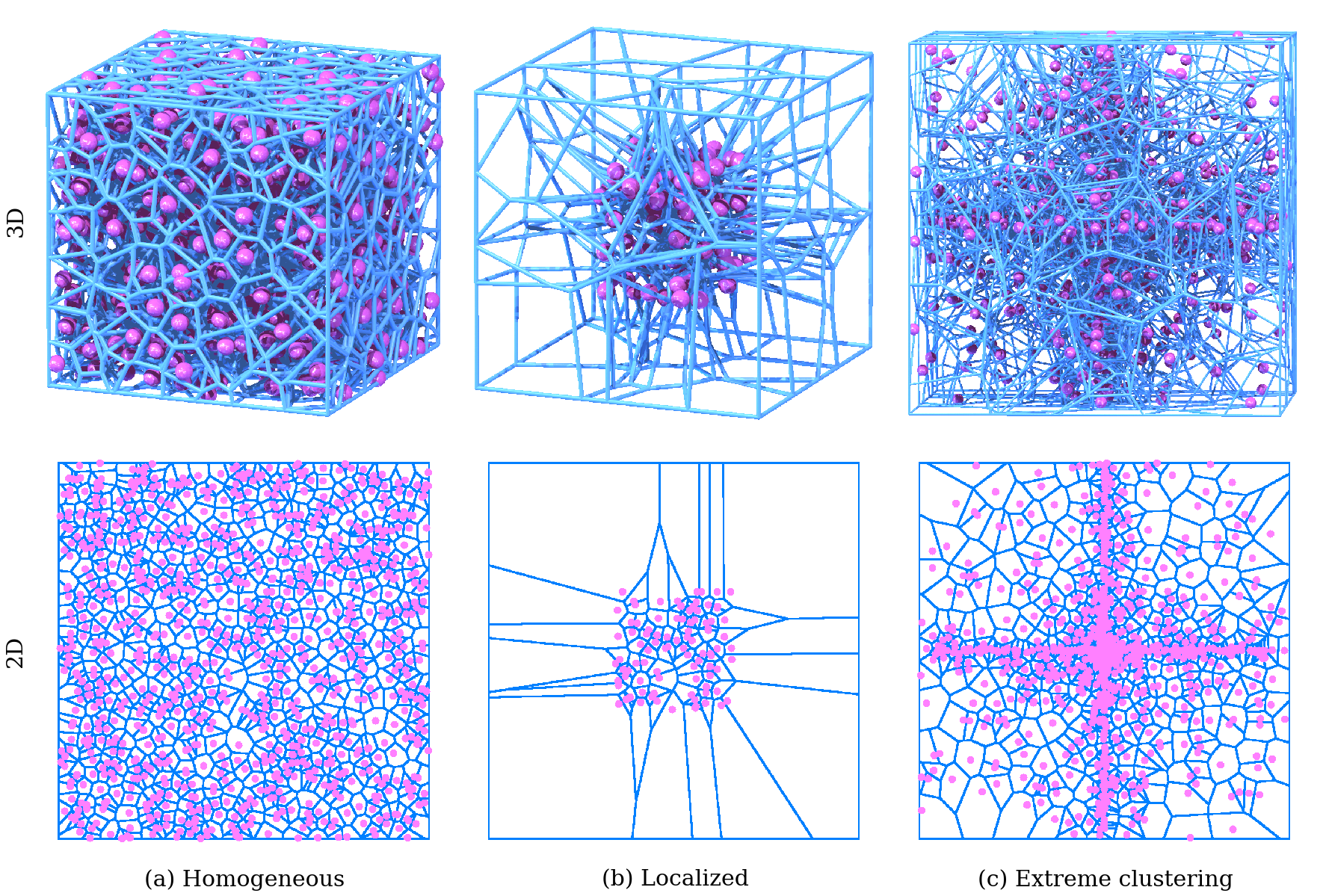}
  \caption{Illustration of three particle distribution cases in both 2D and 3D,
  but scaling down the particle number by $1/10^5$ for ease of visualization.
  (a) Homogeneous and random particle distribution with $1000$ particles. (b)
  Localized and random particle distribution with $95$ particles. (c) Extreme
  clustering particle distribution with $1000$ particles.\label{fig:sec3_cases}}
\end{figure}

We first examine the performance of the multithreaded code for three
representative non-periodic particle arrangements, using both a 2D domain
$[0,1]^2$ and 3D domain $[0,1]^3$. They are:
\begin{enumerate}
  \item \textit{Homogeneous} -- $10^8$ particles randomly and homogeneously
    distributed throughout the domain.
  \item \textit{Localized} -- We define a square of area 0.1 in 2D, or a cube
    of volume 0.1 in 3D, positioned in the center of the domain. We take the
    $10^8$ particles from the previous case, and keep only those within this
    region, resulting in approximately $10^7$ particles in total.
  \item \textit{Extreme clustering} -- We use an uneven distribution clustered
    around the regions $x=\tfrac12$, $y=\tfrac12$, and $z=\tfrac12$ (for 3D
    only). This is achieved by generating each particle coordinate as a
    non-uniformly distributed random variable. In the $x$ direction, the number
    $X$ is chosen to be uniformly distributed over $[0,1]$, and then the
    particle coordinate is set to $x=\tfrac12 + 4(X-\tfrac12)^3$, resulting in
    clustering near $x=\tfrac12$. The same procedure is used in the other
    directions.
\end{enumerate}
Figure~\ref{fig:sec3_cases} shows examples of the three particle arrangements
in both 2D and 3D, but with a reduced number of particles for ease of
visualization. The grid of blocks is set according to the method described in
Sec.~\ref{sec:blocks}, but with specific choices of $\nopt$ are shown in Table
\ref{tbl:sec3_container_size}. For the homogeneous case we use the default
choices of $\nopt$. For the other two cases, we scan over a range of values of
$\nopt$ and choose the one that results in the fastest execution. Since
the structure of these two cases differs considerably from the homogeneous
arrangement, we find that the best values of $\nopt$ are quite different. For
example, for the extreme clustering case we use the much smaller value of
$\nopt=0.5$, since this provides better spatial sorting of the clustered
particles.

\begin{table}
  \begin{center}
  \small
  \begin{tabular}{|p{40mm}|c|c|c|c|}
\hline
\multirow{2}{*}{\parbox{40mm}{\centering Container initialization}}
& \multicolumn{2}{c|}{2D}
& \multicolumn{2}{c|}{3D}
\\ \cline{2-5}
&$\nopt$&$n_x,n_y$&$\nopt$&$n_x,n_y,n_z$
\\ \hline
  (a) Homogeneous&3.4&5,423&4.6&279
\\ \hline
  (b) Localized&30&1,825&49&126
\\ \hline
  (c) Extreme clustering&0.5&14,142&0.5&584
\\ \hline
\end{tabular}
\end{center}\vspace{-0.8em}
\caption{Choices of $\nopt$ and the size of the grid of blocks for the three
different example particle configurations, in 2D and 3D.\label{tbl:sec3_container_size}}
\end{table}

Here, and throughout Sec.~\ref{sec3:parallel performance:main header} all tests
were performed on a Ubuntu Linux computer with dual Intel Xeon E5-2650L v4
processors with 14 low-power cores and a 1.7\,GHz base clock speed. We measure
the time to compute all of the Voronoi cells, without performing any analyses
on them. For each case, we compare the five OpenMP load balancing strategies
introduced in Sec.~\ref{sec:load_balancing}. For the two chunk-based
strategies, we first scan over a range of chunk sizes from $1$ to $100$ with
the 28-thread parallel code, and choose the one with the lowest computation
time. We report this chunk size in our results.

The relevant measures of parallel performance are:
\begin{enumerate}
  \item The wall-clock time $t_p$ required to compute all Voronoi cells using
    $p$ threads.
  \item The parallel efficiency calculated as
    \begin{equation}
      T_e(p) =\frac{t_1}{p \cdot t_p},
      \label{eq:orig_eff}
    \end{equation}
    which measures the effective slowdown from the hypothetically perfect
    parallel scaling.
\end{enumerate}
The computation time and parallel efficiency are affected by Intel's Turbo Boost
technology, which boosts the clock speed based on the number of
active cores. For the E5-2650L v4 chip, the clock speed is boosted by 0.8\,GHz
when a single core is in use, but this is reduced to 0.3\,GHz when six or more
cores are used. This creates a clear signature in the parallel efficiency data,
since the wall-clock times for many threads are increased because the CPU slows
down. We explore this issue in detail in Appendix \ref{appendix:turbo_boost}.
Since we aim to examine the performance of the code independent of hardware
intricacies, we work with the \textit{adjusted parallel efficiency}, which
factors out the Turbo Boost effect. This is computed as
\begin{equation}
  \label{eq:adj_eff}
  T_e^\text{adj}(p) = T_e(p) A(p),
\end{equation}
where
\begin{equation}
  \label{eq:adj_factor}
  A(p)=\frac{\text{average clock speed for $1$ thread}}{\text{average clock speed for $p$ threads}}.
\end{equation}
The details of how the average clock speeds are computed are provided
in Appendix \ref{appendix:turbo_boost}. We account for Turbo Boost in the
parallel efficiency, but we report the original wall-clock computation times,
since they give a reasonable estimate for the Voronoi cell computation in a
real world setting where Turbo Boost is enabled by default.

\subsubsection{Comparison of the three example cases}
\begin{figure}
  \centering
  \includegraphics[width=1\textwidth]{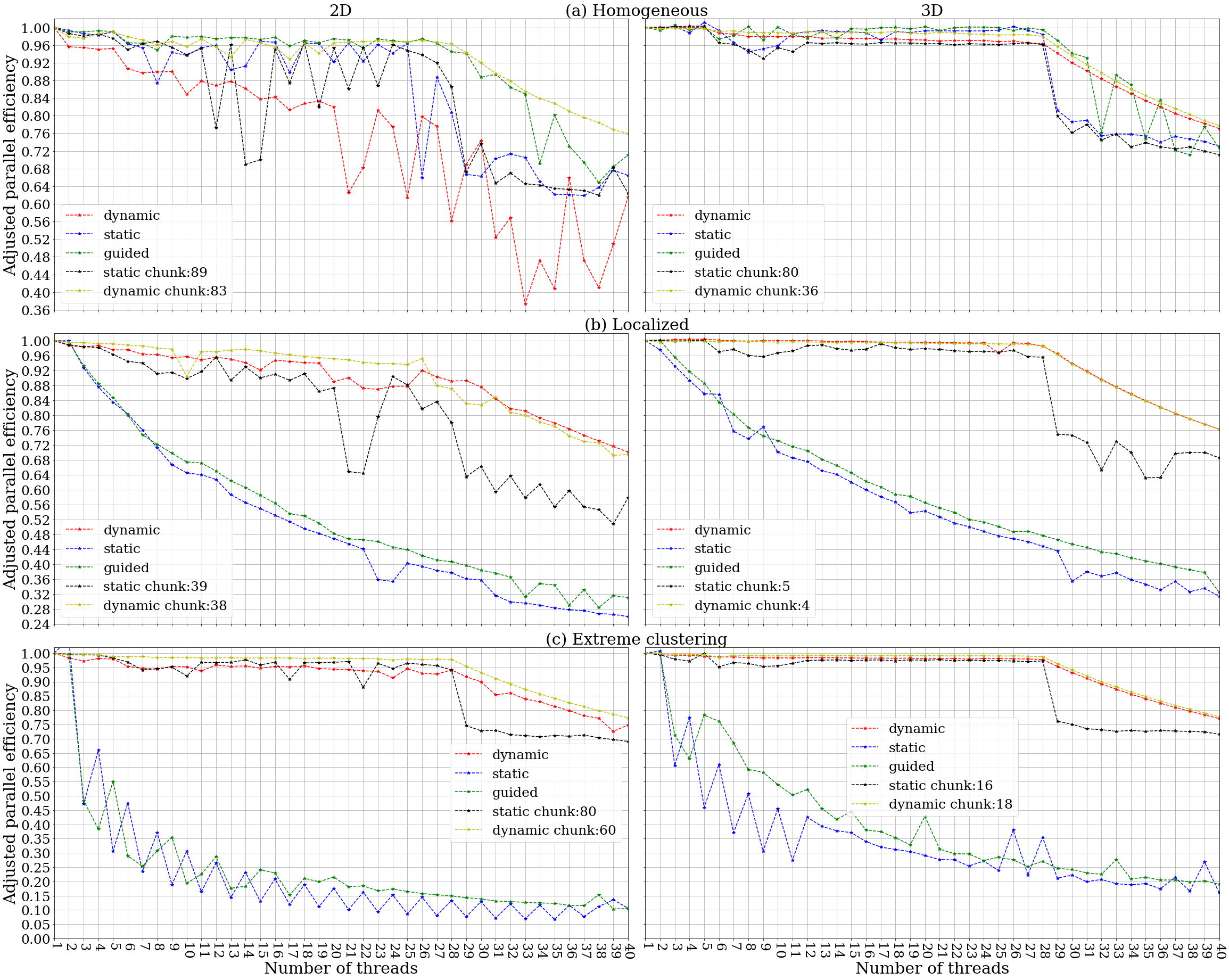}
  \caption{Adjusted parallel efficiency against number of threads of the five load
  balancing strategies, for the three example cases in both 2D and 3D. (a)
  Homogeneous and random particle distribution. (b) Localized and random
  particle distribution. (c) Extreme clustering particle
  distribution.\label{fig:sec3_eff_threads}}
\end{figure}

Figure \ref{fig:sec3_eff_threads} shows the adjusted parallel efficiency of the
five strategies as a function of the number of threads, for the three different
particle arrangements in both 2D and 3D. Since the computer has 28 physical
cores, we find that the parallel efficiency drops substantially beyond 28
threads in all cases, and thus we limit our discussion to the results for the
first 28 threads. Table \ref{tbl:sec3_cases_timing} shows the computation time
using the serial code, and the optimal 28-thread parallel code for the three
cases in both 2D and 3D. For each 28-thread time, we report the parallel
strategy that achieves this optimal result.

\begin{table}
  \begin{center}
  \footnotesize
\begin{tabular}{|p{34mm}|c|c|c|c|c|c|}
\hline
\multirow{2}{*}{\parbox{34mm}{\centering Wall-clock time (s)}}
& \multicolumn{3}{c|}{2D}
& \multicolumn{3}{c|}{3D}
\\ \cline{2-7}
&1 thread&28 threads&Opt.\@ strategy&1 thread&28 threads&Opt.\@ strategy
\\ \hline
  (a) Homogeneous&150.53&7.03&(guided)&2099.72&93.34&(guided)
\\ \hline
  (b) Localized&39.88&2.03&(dynamic,38)&589.62&26.85&(dynamic)
\\ \hline
  (c) Extreme clustering&1030.12&45.45&(dynamic,60)&10368.70&464.96&(dynamic,18)
\\ \hline
\end{tabular}
\end{center}\vspace{-0.8em}
\caption{Voronoi computation time with the serial code and the $28$-thread parallel code for the three example cases in both 2D and 3D, using each case's optimal strategy.\label{tbl:sec3_cases_timing}}
\end{table}

\paragraph{Comparison of different particle arrangements}\mbox{}
Figure \ref{fig:sec3_eff_threads} shows that for the homogeneous particle
distribution the parallel efficiencies are similar among the strategies, and
all are in the high range above 80\% in 2D and 95\% in 3D. In comparison, for
the localized particle distribution, the parallel efficiencies are very
different, with \co{schedule(guided)} and \co{schedule(static)} performing
significantly worse than the others. This large difference is also true for the
extreme clustering case. For a container with random and homogeneous particle
arrangement, we expect that the computation time for each Voronoi cell to be
similar, as the Voronoi cells are of similar size. Therefore, particles have
similar workloads. No matter what strategies we use to distribute particles to
threads, the threads have similar workloads as well, thus resulting in the
similar performance among the five strategies.

However, for the random and localized distribution of particles, where the
sizes of Voronoi cells are only restricted by the container, many particles in
the outer layer have extended Voronoi cells that take much longer to compute as
described in Sec.~\ref{sec:voro_compute}. The imbalance in the workload causes
\co{schedule(static)} and \co{schedule(guided)} to perform substantially worse,
with parallel efficiencies dropping below 50\%. \co{schedule(dynamic)} performs
much better, achieving efficiencies above 90\% in 2D and 95\% in 3D. In
addition, both \co{schedule(dynamic,chunk)} and \co{schedule(static,chunk)} can
have similar parallel performance to \co{schedule(dynamic)}, if optimal chunk
sizes are chosen. Table \ref{tbl:sec3_cases_timing} shows that the total
computation times for the localized case are less than the homogeneous case,
although the localized case only has a tenth of the particles of the homogeneous
case. The localized case takes considerably longer per particle.

For the extreme clustering case, there are many more particles concentrating in
the grid blocks in the domain center. The Voronoi computation for these
particles is substantially more expensive than the Voronoi computation for
particles in less dense regions. When \vpp{} computes Voronoi cell of a
particle, it will loop through all neighboring particles in the nearby grid
blocks (Sec.~\ref{sec:voro_compute}), and thus more particles will be
considered. In addition, the uneven distribution creates imbalanced workload
since the particles in the center will take longer to compute than those in the
periphery. Therefore, similar to the localized case, we see that
\co{schedule(dynamic)} has huge benefits and performs much better than
\co{schedule(static)} and \co{schedule(guided)}. Also,
\co{schedule(static,chunk)} and \co{schedule(dynamic,chunk)} are able to
perform as well as \co{schedule(dynamic)}, if optimal chunk sizes are chosen.
Table \ref{tbl:sec3_cases_timing} shows that the total time for the extreme
clustering case is considerably longer than for the homogeneous case.

\paragraph{Comparison of 2D and 3D Voronoi computation}\mbox{}
3D Voronoi cells are much more complicated than 2D Voronoi cells. A 2D Voronoi
cell is a simple polygon, where every vertex is connected to two others. By
contrast, representing a 3D Voronoi cell requires an edge table with varying
connectivity between the vertices. This difference is clear in the timing data
for the homogeneous case in Table \ref{tbl:sec3_cases_timing}: the serial code
computes 664000 cells per second in 2D, but only 47600 cells per second in 3D.

Since each 3D Voronoi cell computation involves a larger amount of work, the
code can achieve higher threading efficiency in 3D, because comparatively less
time is lost to overhead in parallelizing the loop. In
Fig.~\ref{fig:sec3_eff_threads} for the homogeneous case with
\co{schedule(dynamic)}, the 2D computations only achieve efficiencies of
$\approx 75\%$ for large thread counts, whereas the 3D computations achieve
efficiencies over 95\%. The data in Fig.~\ref{fig:sec3_eff_threads} shows that
the \co{schedule(dynamic,chunk)} strategy substantially improves the efficiency
of the 2D computations, since proportion of time spent on thread assignment
will be reduced. With this strategy, the 2D computations can achieve thread
efficiencies approaching 95\%, as in the 3D case.

\subsubsection{Code performance for different system sizes}
We now examine the performance of the code for a random homogeneous particle
system when the total number of particles $N$ is varied. For each case, the
grid of blocks is again set according to the method described in
Sec.~\ref{sec:blocks}, and the values of $\nopt$ remain $3.4$ for 2D and $4.6$
for 3D. We use the \co{schedule(guided)} strategy, which was previously shown
to achieve very good parallel efficiency when $N=10^8$, in both 2D and 3D
(Fig.~\ref{fig:sec3_eff_threads}). We compute parallel efficiencies by
comparing the performance of a single thread to the performance of 28 threads.

\begin{figure}
  \centering
  \includegraphics[width=0.8\textwidth]{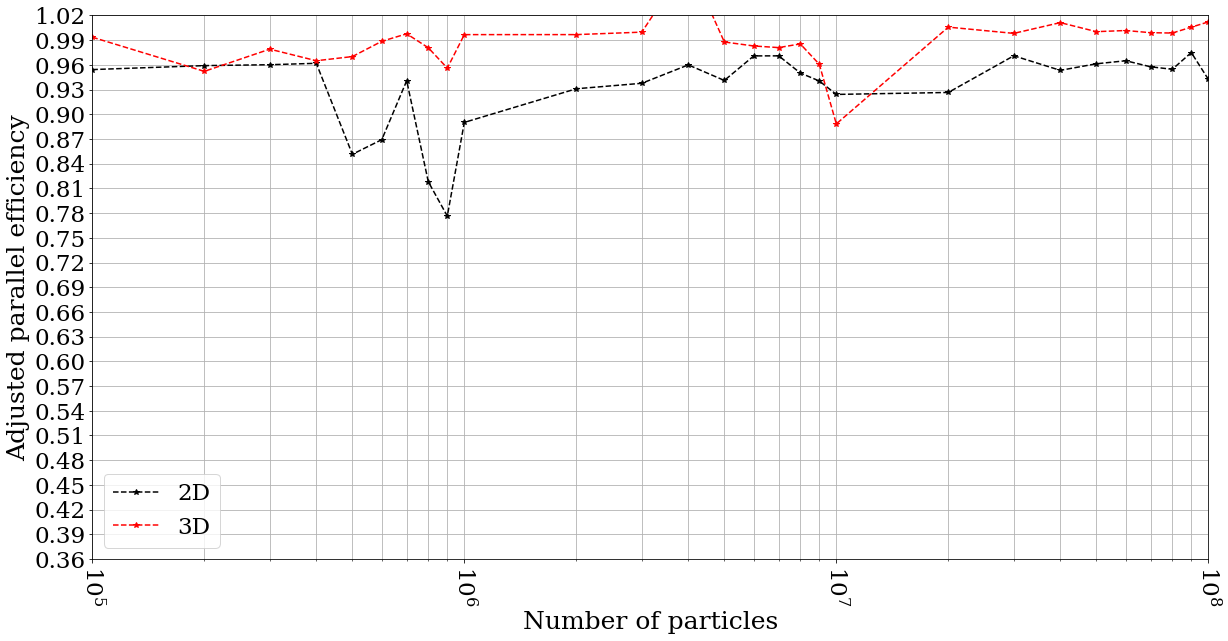}
  \caption{Adjusted parallel efficiency as function of the number of particles.
  The data obtained is from the homogeneous particle distribution case in both
  2D and 3D, both using \co{schedule(guided)}. While parallel efficiency must
  theoretically be less than or equal to one, small random fluctations in
  timing cause several data points in this graph to exceed this value.
  \label{fig:sec3_eff_particles}}
\end{figure}

As shown in Fig.~\ref{fig:sec3_eff_particles}, the parallel efficiencies remain
roughly constant over the range from $N=10^5$ to $N=10^8$. The Voronoi cell
computations are well suited to parallel computation, and even for $N=10^5$
there is sufficient work for the particles to be parallelized effectively. As
expected, the parallel efficiencies seen in 2D are lower than those in 3D.
Some small fluctuations in parallel efficiency are observed in
Fig.~\ref{fig:sec3_eff_particles}, particularly for lower particle counts in 2D.
Because the wall clock times for these computations are shorter, they are more
susceptible to small variations in computation time.

\subsubsection{Effects of clipping extended Voronoi cells}
\begin{figure}
  \centering
  \vspace{-1em}\includegraphics[width=0.4\textwidth]{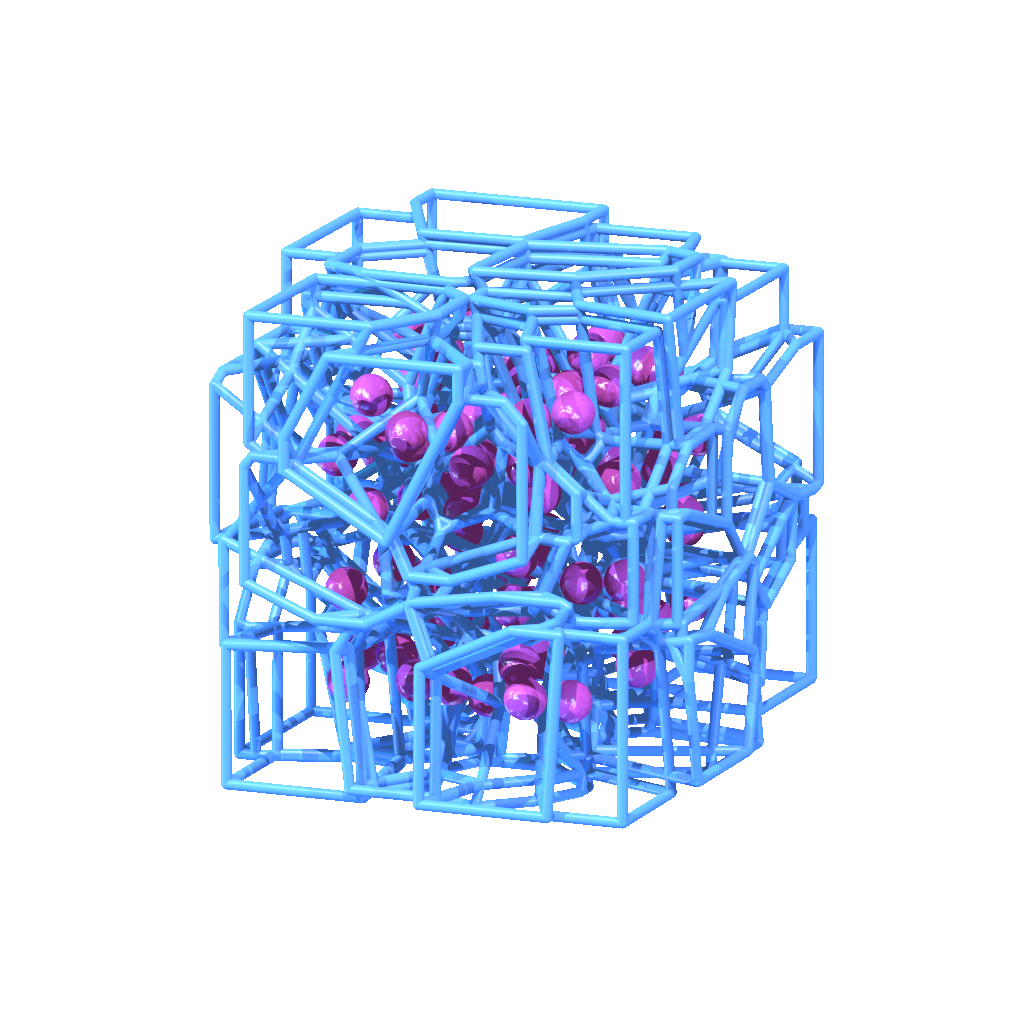}\vspace{-1em}
  \caption{Illustration of the 3D localized particle distribution test from
  Fig.~\ref{fig:sec3_cases}, but where the Voronoi cells are each clipped
  by a bounding cube of side length $h$ as described in the text. Here, the
  total number of particles has been scaled down by a factor of $10^5$ from the
  computational test for ease of visualization. Correspondingly, the clipping
  length $h$ has been scaled up by a factor of
  $10^{5/3}$.\label{fig:sec3_cases_localized_bound}}
\end{figure}

To illustrate further the effect of extended Voronoi cells on the computation
cost, we examine another case where the Voronoi cells are clipped by a bounding
volume. This is demonstrated in Fig.~\ref{fig:sec3_cases_localized_bound} where
the Voronoi cells for the 3D localized arrangement of Fig.~\ref{fig:sec3_cases}
are clipped by a bounding cube centered on each particle. Clipped Voronoi cells
are useful in some practical situations~\cite{murphy14,honeyager16}, such as
when particles have a finite interaction range, and extended Voronoi cells
provide no practical information.

\vpp{} can clip Voronoi cells by any convex polyhedron centered on each
particle. This is achieved by starting each Voronoi cell computation using a
fixed shape, instead of making it fill the computational domain as described in
Sec.~\ref{sec:using_cell}. Here, we consider the 3D test using the localized
particle arrangement shown in Fig.~\ref{fig:sec3_cases}, with the domain
$[0,1]^3$. This configuration was made by generating $N'=10^8$ candidate
particle positions at random, and then retaining the $N\approx 10^7$ in a
central cube covering $10\%$ of the domain volume. We clip each Voronoi cell
using a cube with side length $h=1.5\sqrt{3}d_0$, where $d_0=(N')^{-1/3}$
represents a typical inter-particle separation length in the container.
Figure \ref{fig:sec3_cases_localized_bound} shows an illustration of the
clipped Voronoi cells.

\begin{figure}
  \centering
  \includegraphics[width=0.8\textwidth]{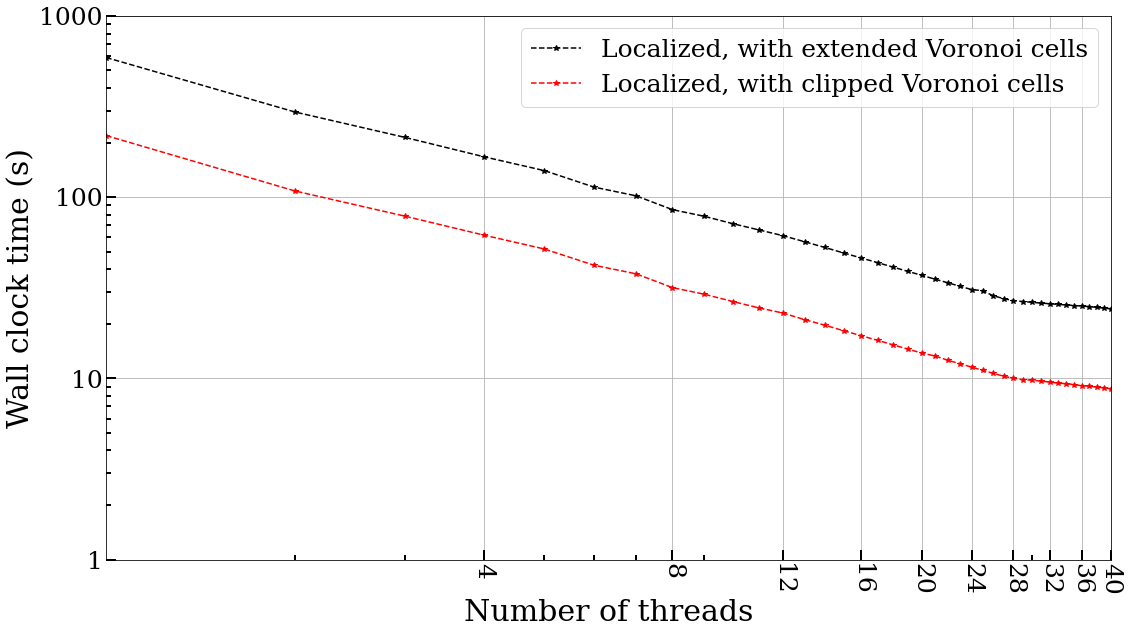}
  \caption{Computation time using different numbers of threads for the 3D
  localized particle distribution test using (a) the original Voronoi
  tessellation with many extended Voronoi cells, and (b) clipping each
  Voronoi cell by a bounding cube of side length $h$ as described in the text.
  In both cases, the optimal strategy of \co{schedule(dynamic)} was
  used.\label{fig:sec3_time_loc_wwo_walls}}
\end{figure}

As shown in Fig.~\ref{fig:sec3_time_loc_wwo_walls}, for the same localized
particles in the container, using \co{schedule(dynamic)}, the
computation time is significantly lower for clipped Voronoi cells. For the
original localized case of extended Voronoi cells, the computation time is
589.62\,s for one thread and 26.85\,s when using 28 threads. In comparison,
when the Voronoi cells are clipped, the computation time decreases to 218.63\,s
with one thread and 10.01\,s with 28 threads. Therefore, compared to the
extended Voronoi cells, the computational costs for clipped Voronoi cells are
significantly lower, and in both serial and parallel cases, there is a
three-fold speedup in computation time. These timing results will depend on
the size of the clipping region, with larger $h$ requiring more time, and
smaller $h$ requiring less time. With the clipped cells, the size of the
search space in the Voronoi cell construction (Sec.~\ref{sec:voro_compute}) is
substantially reduced, improving the performance. Furthermore, using clipped
cells removes the large disparities in computation time between different
Voronoi cells, so the \co{schedule(static)} parallelization strategy will give
good performance.

\subsection{Performance of the parallel insertion routine}
\label{sec:parallel_put_start}
We now test the performance of the parallel insertion routine that was
described in Sec.~\ref{sec:parallel_put}. We test in 2D and 3D using the
domains $[0,1]^2$ and $[0,1]^3$, respectively, with $10^8$ homogeneous random
particle positions. We use the default values of $\nopt=3.4$ and $\nopt=4.6$
for the 2D and 3D calculations, respectively. In each case we use an initial
assignment of \co{mem[k]} equal to 8. The total number of particles in each
block is well-approximated by a Poisson distribution, $\text{Pois}(\nopt)$. This
results in 0.83\% and 4.51\% of blocks exceeding the initial memory allocation
in 2D and 3D, respectively. We expect 0.36\% and 1.66\% of the total particles
will be placed into the overflow buffer in 2D and 3D, respectively.

We note that the performance of parallel insertion will depend substantially on
many factors. Due to the small difference in $\nopt$ between 2D and 3D, this
results in a 4.6-fold difference in the number of particles going into the
overflow buffer. Since the overflow buffer is processed serially at the end of
the insertion, this will affect parallel efficiency. This will also depend upon
the particle configuration; for example, molecular dynamics simulations
often employ a repulsive potential at a short, well-defined length scale
(\textit{e.g.} the Lennard-Jones potential~\cite{lennard-jones31}). This causes
particles to be spatially distributed more evenly than uniform random samples,
so that the overflow buffer may not be needed. In contrast, particle
configurations like the extreme clustering case of Fig.~\ref{fig:sec3_cases}
may have a large fraction of particles going to the overflow buffer.

In many practical uses of \vpp{}, the Voronoi tessellation may be computed at
multiple time points in a simulation, where the particle positions only vary
slightly between successive snapshots. To handle this scenario, \vpp{} has the
ability to re-use the container data structure, by clearing the particles for
inserting a new configuration. This is achieved by zeroing out the block-based
particle counters \co{co[k]}, but leaving any extended memory allocation
\co{mem[k]} in place. Thus, assuming only small movements in particles, the
previously extended block memory should be well-matched to the new particle
configuration and only a small number of particles will need to be placed
in the overflow buffer.

To illustrate this and isolate the effect of the overflow buffer, we perform a
sequence of two parallel insertion tests, in both 2D and 3D. We generate a list
of $N$ uniformly distributed particle positions, and call the
\co{add\_parallel(...)} function to add these particles into the container. On
the first time, some particles will be placed into the overflow buffer, and
will then be reconciled and store serially via the
\co{put\_reconcile\_overflow()} function. This will result in \co{mem[k]} being
adjusted to accommodate the additional particles. We then clear the container
and make another \co{add\_parallel(...)} call to reinsert the particles. Since
the block memory was already extended to accommodate this particle
configuration, zero particles will be placed into the overflow buffer on this
second test. Each test was performed ten times with between $1$ and $28$
threads, and the average wall-clock time was recorded to calculate the average
parallel efficiency. Since the parallel insertion routine involves both
parallel and serial components, the changing clock frequency complicates
the performance analysis. We therefore turn off Turbo Boost, so
that all cores run at the same clock frequency regardless of the number of
active threads. The average wall-clock time for the serial code and the
28-thread parallel code in both 2D and 3D are reported in
Table.~\ref{tbl:sec3_put_timing}.

\subsubsection{Efficiency against number of threads}
\label{sec:parallel_put_efficiency}
\begin{figure}
  \centering
  \includegraphics[width=1\textwidth]{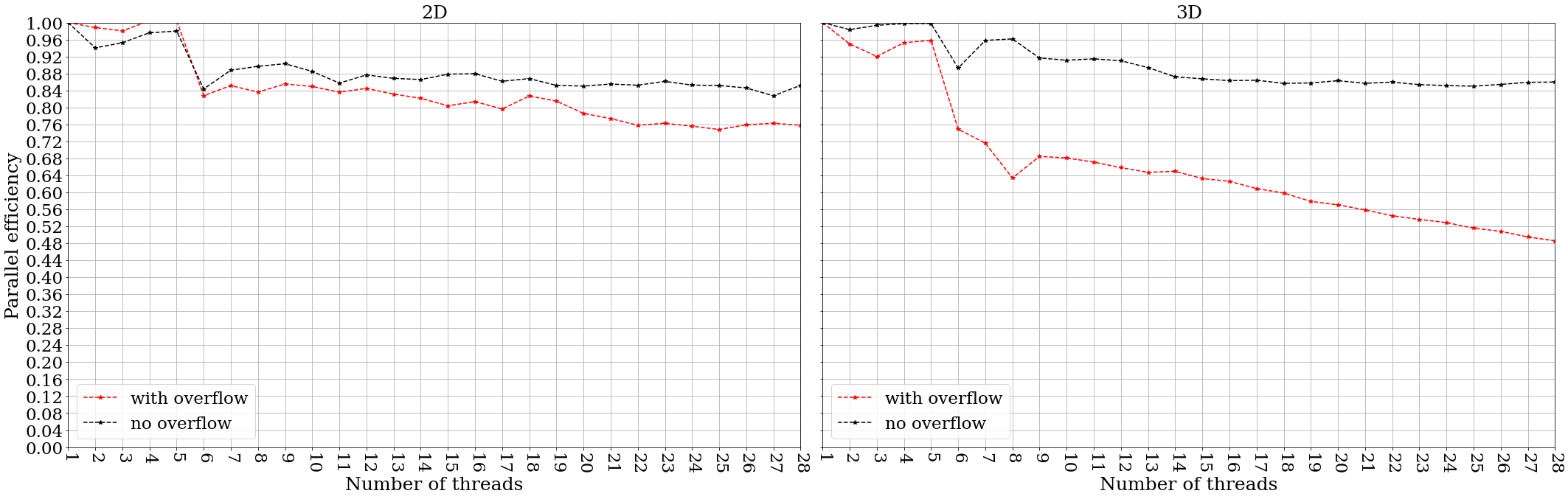}
  \caption{Average parallel efficiency for the parallel particle insertion test of $10^8$ particles using a varying number of threads. For each case, the insertion is performed once, when the overflow buffer will be used. The container is cleared and the insertion is repeated. Since block memory was extended during the first test, the overflow buffer will not be used on the second insertion. Each test was performed ten times, and the average wall-clock times were recorded to calculate the average parallel efficiency. While parallel efficiency must theoretically
be less than or equal to one, small random fluctations in timing cause several data points in the 2D graph (left) to exceed this value.\label{fig:sec3_put_eff_thread}}
\end{figure}

Figure \ref{fig:sec3_put_eff_thread} show the results of parallel insertion in
2D and 3D using $N=10^8$ particles with a varying number of threads. On the
first time, when the overflow buffer is used, the parallel efficiency in 2D is
approximately 76\% using 28 threads. In 3D, parallel efficiency is lower, and
gradually decreases to 29\% when 28 threads are used.

On the second time, without using the overflow buffer, the $28$-thread parallel
efficiency is approximately 85\% in both 2D and 3D. These results indicate how
the overflow buffer substantially affects the efficiency. The worse parallel
performance in 3D is due to more particles being placed into the overflow
buffer. The number of particles being placed into the overflow buffer averaged
over ten tests are $356,044$ and $1,675,330$ for 2D and 3D, respectively,
which are consistent with the aforementioned calculations using the Poisson
distribution.

\begin{table}
\begin{tabular}{|p{75mm}|c|c|c|c|}
\hline
\multirow{2}{*}{\parbox{75mm}{\centering Wall-clock time (s), \co{add\_parallel(\dots)}}}
& \multicolumn{2}{c|}{2D}
& \multicolumn{2}{c|}{3D}
\\ \cline{2-5}
 &1 thread&28 threads&1 thread&28 threads
\\ \hline
  With overflow buffer&72.76&3.43&72.65&5.34
\\ \hline
  No overflow buffer&72.37&3.03&73.65&3.06
\\ \hline
\end{tabular}
\caption{Average computation time for the parallel particle insertion test. For each case, the insertion is performed once, when the overflow buffer will be used. The container is cleared and the insertion is repeated. Since block memory was extended during the first test, the overflow buffer will not be used on the second insertion. Each test was performed ten times, and the average wall-clock times are shown for the serial code and $28$-thread parallel code in both 2D and 3D.\label{tbl:sec3_put_timing}}
\end{table}

\subsubsection{Efficiency against number of particles}
\begin{figure}
  \centering
  \includegraphics[width=1\textwidth]{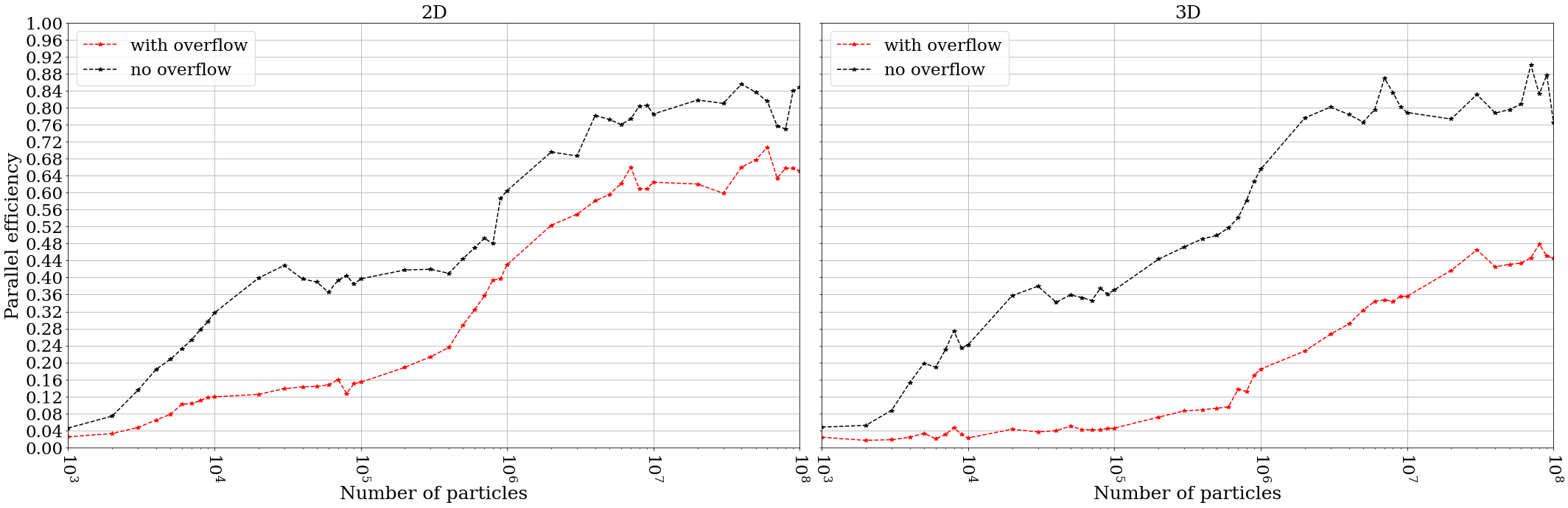}
  \caption{Average parallel efficiency for the parallel particle insertion test of a varying number of particles, using 28 threads. For each case, the insertion is performed once, when the overflow buffer will be used. The container is cleared and the insertion is repeated. Since block memory was extended during the first test, the overflow buffer will not be used on the second insertion. Each test was performed ten times, and the average wall-clock times were recorded to calculate the average parallel efficiency.\label{fig:sec3_put_effN}}
\end{figure}

Figure \ref{fig:sec3_put_effN} shows the parallel efficiency using 28 threads
as number of particles $N$ is varied. For both 2D and 3D, we observe a clear
trend of increasing parallel efficiency for larger $N$, demonstrating that
parallel insertion becomes more effective for larger particle configurations.
We also see that with overflow buffer, the 3D case has worse parallel
efficiency. Without overflow buffer, the parallel efficiencies are comparable
in 2D and 3D, consistent with the results of Sec.~\ref{sec:parallel_put_efficiency}.

\section{Application: Topological analysis using \votop{}}
\label{sec:vorotop}
Voronoi tessellations are commonly used, among other applications, in the classification
of local, structural features of atomistic data sets \cite{lazar15,brostow1998voronoi,poupon2004voronoi,bernaschi17}.  Computational material scientists and
physicists, for example, are often interested in identifying localized defects in materials \cite{lazar15}, or
else in characterizing the structure of nominally disordered systems such as liquids \cite{shih1994voronoi,starr2002we},
granular packings \cite{rycroft06a,shahinpoor1980statistical}, and soft glassy materials \cite{derzsi2017fluidization,lulli2018metastability}.  Voronoi cells provide a proxy for describing local
arrangements through consideration of their geometric and combinatorial properties \cite{lazar2022voronoi,klatt2014characterization,torquato2010jammed}.

\begin{figure}
  \centering
  \includegraphics[width=0.25\textwidth]{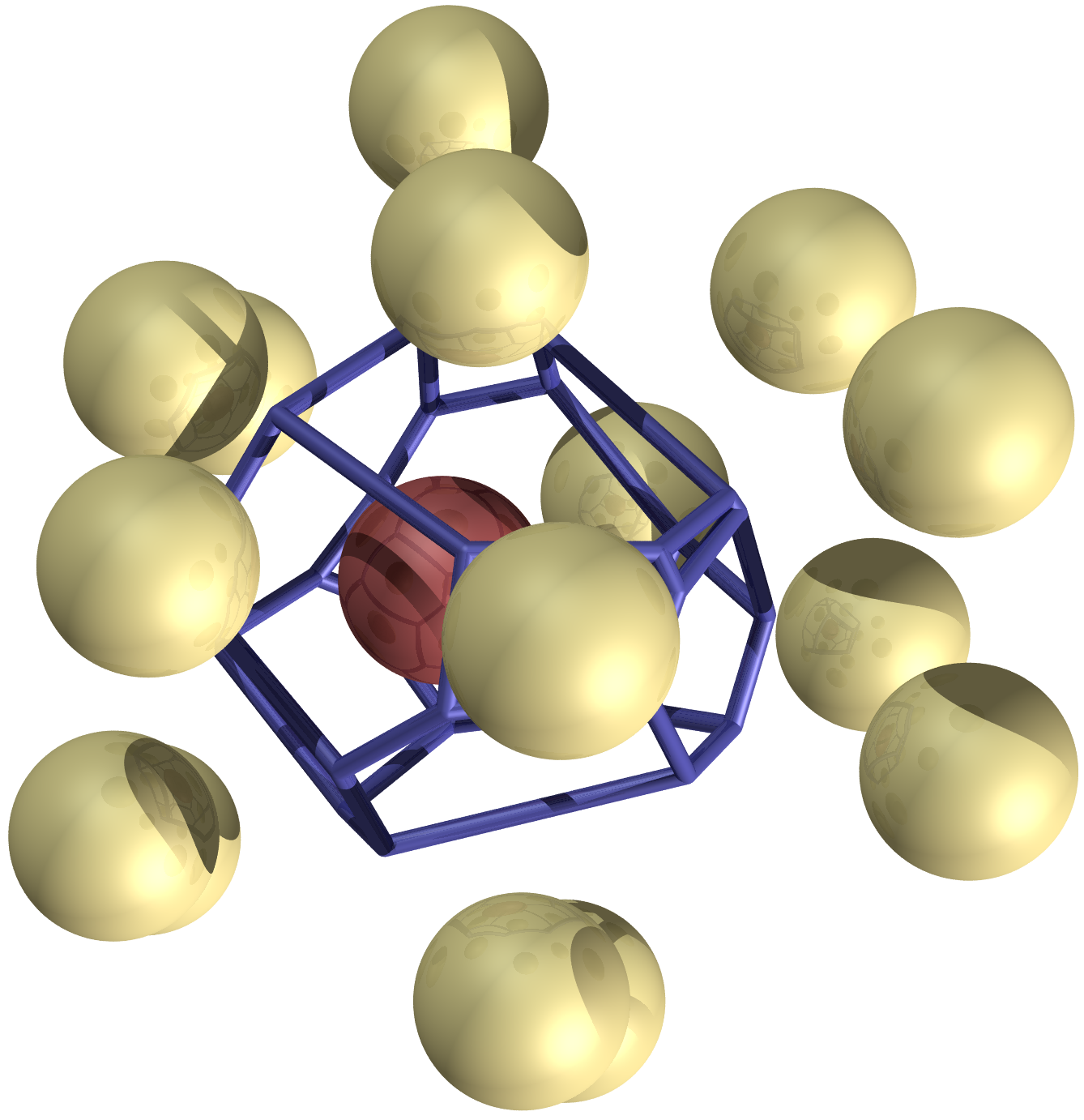}
  \caption{A central red atom and its Voronoi cell, surrounded by gold-colored neighboring atoms.\label{fig:vcell}}
\end{figure}

Voronoi topology provides a robust method of classifying local
structure~\cite{lazar15,lazar12}.  In particular, the types of faces of a
Voronoi cell and the manner in which they are arranged describe the manner in
which neighbors are arranged relative to a central particle and to one another.
Figure \ref{fig:vcell} illustrates a central red atom and its Voronoi cell,
surrounded by gold-colored neighboring atoms.  Each face of the Voronoi cell
corresponds to a unique neighboring particle, and so the total number of faces
can be considered as a count of neighbors.  Furthermore, the number of edges of
a face counts the number of common neighbors between the central particle and
the associated neighbor.  A five-sided face, for example, indicates that the
central particle shares five neighbors in common with the associated particle.
In finite-temperature systems, particles are located in ``general position''
and small perturbations of the coordinates do not change features such as
numbers of faces and edges, making them particularly useful for analyzing noisy
data.  Voronoi topology is thus particularly well-suited for studying
high-temperature crystalline systems, where conventional methods such as
centrosymmetry and common-neighbor analysis are typically
ineffective~\cite{stukowski12}.

Fully characterizing the topological structure of a Voronoi cell can be done
efficiently using an algorithm of Weinberg~\cite{weinberg66} designed to
determine whether two planar graphs are isomorphic.  Although the general
graph-isomorphism problem is not known to be solvable in polynomial time, the
edge graphs associated with Voronoi cells, as convex polyhedra, are always
3-connected and planar \cite{steinitz1922polyeder}.  This enables the use of
Weinberg's graph-tracing algorithm to provide a systematic description of
Voronoi cell topologies.  In particular, Weinberg's algorithm traces the edge
network of a graph to produce a unique ``code''; two Voronoi cells are
topologically equivalent if and only if their codes are identical.  This
algorithm takes as input a representation of a planar graph and outputs a code
whose length is twice the number of edges of the graph.  Constructing each code
is done in $O(n^2)$ time, where $n$ is the number of faces of the Voronoi cell.
Voronoi cells in uniformly distributed particle systems typically have a narrow
range of faces (between 10 and 20), but this number can grow for non-uniformly
distributed ones.

Calculating the Voronoi cell topology for large particle systems is a good test
case for a multi-threaded version of \vpp{} because the Weinberg code for each
cell is computed individually after the Voronoi cell itself has been
constructed.  A program called \votop, developed by Lazar, uses \vpp{} to
calculate the Voronoi cells and then implements Weinberg's algorithm to produce
a code for each particle, enabling further structural analysis~\cite{lazar17}.
When limited to running with a single-thread, the ability to analyze large
systems is limited.  While smaller systems can be analyzed separately in
parallel using batch runs, this approach is ill-suited for large systems where
memory constraints may limit the number of systems that can be computed
simultaneously.  We have thus incorporated the multithreaded version of \vpp{}
into an updated version of \votop{} to take advantage of its parallel
architecture.

To test the efficiency of the multithreaded \vpp{} in \votop, we considered
systems with between 100,000 particles and 102.4 million particles.  These
systems were analyzed on a single computer with two Intel Xeon Gold 6240 CPUs
running at 2.60\,GHz.  Each of the two processors has 18 cores, allowing us to
run up to 36 threads while limiting each core to a single thread.  Running
\votop{} and \vpp{} with more threads than physical cores can still reduce
total runtime, even while decreasing overall parallel efficiency.  Running
\votop{} and \vpp{} with more than 72 threads, however, led to decreased total
performance, as the overhead cost associated with running more than two threads
per core outweighed gains from multithreading.  To better understand the impact
of the parallel version of \vpp{} on this topological analysis, no further
computations, such as structure classification, were performed.

As in Sec.~\ref{sec:parallel_put_start}, we turned off Turbo Boost for this
test. This reduces total performance but enables a simpler comparison of
efficiency of the parallel versions of \vpp{} and \votop. Furthermore, we only
considered the \co{schedule(dynamic)} strategy.  We constructed systems with
each given number of particles, distributed randomly and uniformly in the unit
cube with periodic boundary conditions.
\begin{figure}
  \centering
  \includegraphics[width=0.75\textwidth]{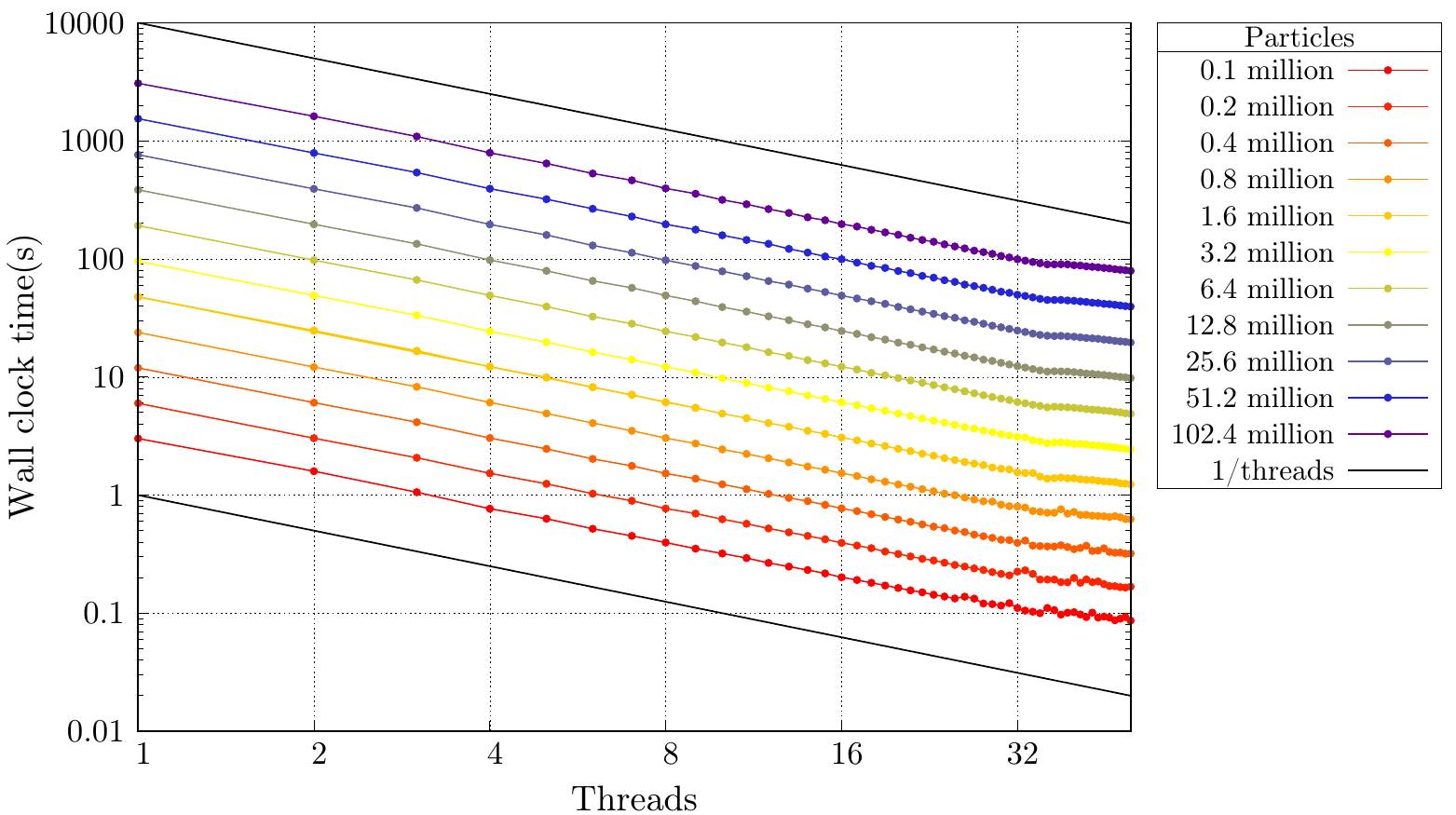}\vspace{2mm}
  \caption{Computation time against number of threads for the different systems.  The two guidelines are graphs of functions proportional to $1/$threads.
  ~\label{fig:vorotop_walltime}}
\end{figure}
Each system was analyzed ten times with between 1 and 50 threads, and the
average wall-clock runtime was recorded for computation of the Voronoi cells
and the computation of their Weinberg codes.

Figure \ref{fig:vorotop_walltime} shows the average wall-clock time for each
system as a function of number of threads. It can immediately be seen that the
total average wall-clock time decreases with additional cores, as expected.
\begin{figure}
  \centering
  \includegraphics[width=0.75\textwidth]{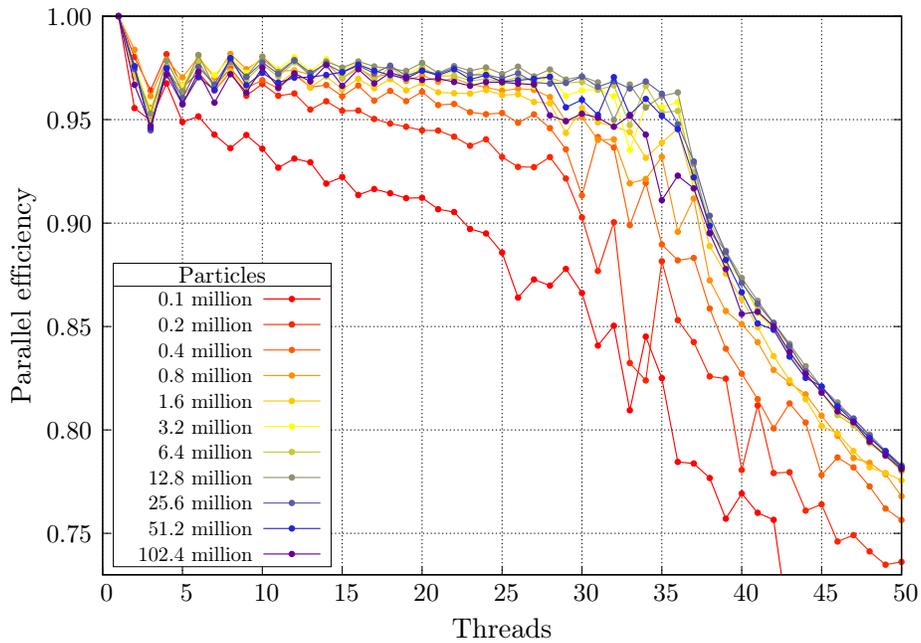}
  \caption{Parallel efficiency as a function of threads for systems with different numbers or particles.  For systems with at least half a million particles, \votop{} runs at over 95\% parallel efficiency when the number of threads is at most the number of physical cores.~\label{fig:vorotop_efficiency}}
\end{figure}
Figure \ref{fig:vorotop_efficiency} shows the unadjusted parallel efficiency
$T_e$ computed via Eq.~\eqref{eq:orig_eff} for different systems sizes and
numbers of threads.  Parallel efficiency is over 95\% when the number of
threads is at most equal to the number of physical cores, in this case 36.
\begin{figure}
  \centering
  \includegraphics[width=0.75\textwidth]{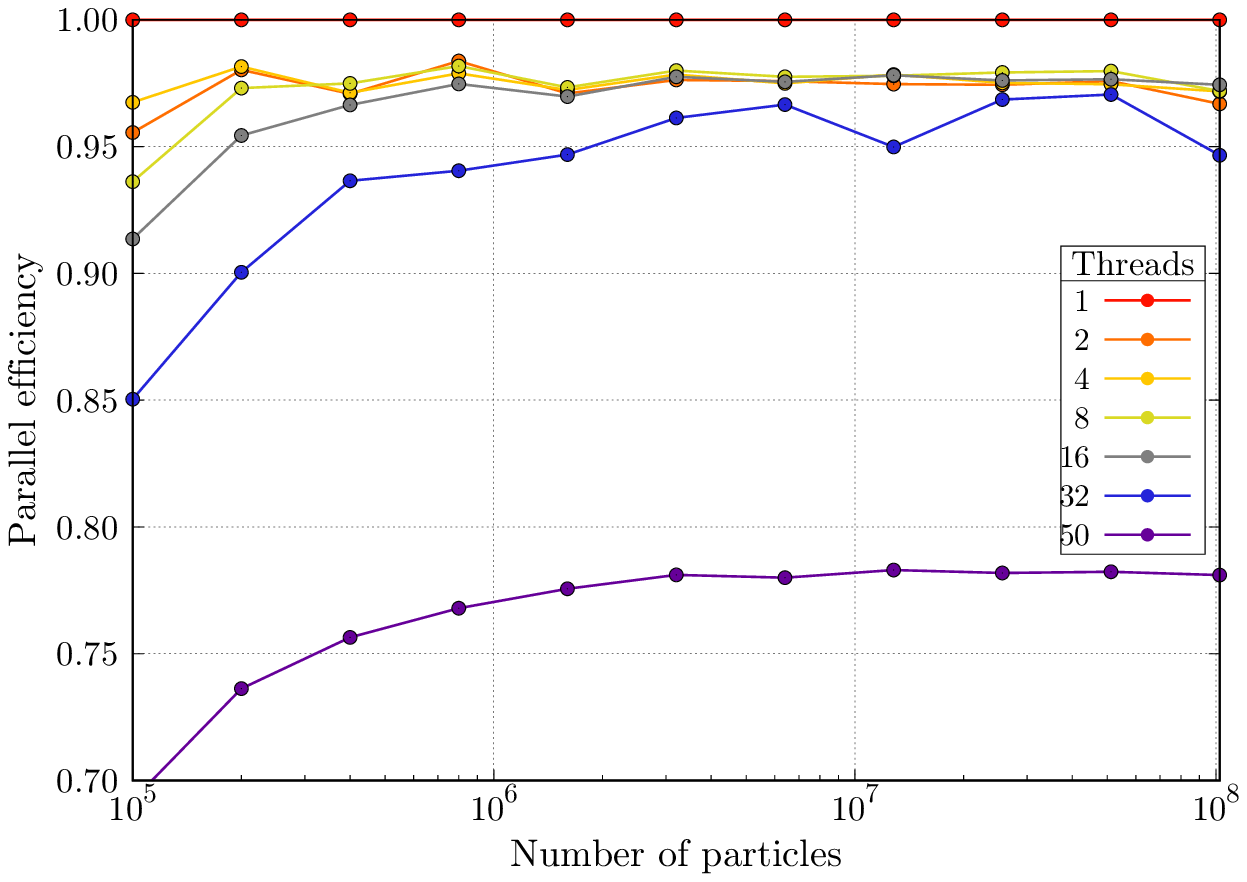}
  \caption{Parallel efficiency as a function of number of particles for different numbers of threads.  For systems with at least half a million particles, \votop{} runs at over 95\% parallel efficiency when the number of threads is at most the number of physical cores.~\label{fig:vorotop_particles_vs_efficiency}}
\end{figure}
Figure \ref{fig:vorotop_particles_vs_efficiency} illustrates the parallel
efficiency as a function of system size, for different numbers of threads. This
efficiency is close to 95\% in systems with at least half a million particles.
Voronoi topology analysis through \votop{} thus provides an example in which
implementation of a multithreaded version of \vpp{} facilitates a significant
speedup in practical applications.

\section{Conclusion}
In this paper we have introduced a multithreaded extension to the \vpp{}
library using the OpenMP standard. Since \vpp{} constructs each Voronoi cell
individually, it is ideally suited for multithreaded computations, and we
demonstrate parallel efficiencies above 95\% across a range of different
examples in 2D and 3D. Our extension to \vpp{} is designed to follow standard
OpenMP programming methods, making it straightforward for users to incorporate
into their own programs. In Sec.~\ref{sec:vorotop}, we presented an example of
this using the \votop{} software package, where both Voronoi cell calculations
and topology analysis were performed within an OpenMP loop, again resulting in
efficiencies above 95\%.

There are a number of extensions to consider. First, as described in
Sec.~\ref{sec:container_2d}, \vpp{} is based upon dividing the computational
domain into a regular grid of blocks. This simple data structure is efficient
to compute on. It is also well-suited to roughly even particle distributions,
since each block is responsible for a similar number of particles. One of the
primary usages of \vpp{} has been in analysis of molecular dynamics (MD) or
discrete-element method (DEM) simulation, where particles often have a fixed
interaction length, and therefore tend to have roughly even distributions,
leading to good performance. However, there are other scenarios, such as
computing the Voronoi tessellation for gravitationally interacting particles,
where particles may be heavily clustered in small parts of the domain. In this
case, the particles are unevenly distributed, resulting in worse performance.
This was demonstrated in the extreme clustering example, which ran
substantially slower than the homogeneous example despite having the same number
of particles (Table \ref{tbl:sec3_cases_timing}). One possibility to improve
this would be to sort the particles into a structure like a $k$-d
tree~\cite{bentley75} or quad/octree~\cite{finkel74}, which would allow
clustered regions of particles to be adaptively refined. This would entail a
more complicated search through the blocks when computing a Voronoi cell, since
the blocks would no longer be of uniform size. Applying a restriction in the
tree construction (\textit{e.g.}~using a graded grid~\cite{chen07,theillard12})
may help simplify the implementation and reduce the number of cases that need
to be considered. Since the tree data structure would even out the particles
among the blocks, it would likely improve the parallel efficiency of the OpenMP
implementation.

Another possible extension is to explore distributed memory calculations, which
would allow the library to handle even larger particle systems, such as those
generated by supercomputing facilities. Furthermore, many codes that are
designed for supercomputers already use distributed parallel programming, such
as via the Message Passing Interface (MPI). Thus if \vpp{} could work in a
distributed parallel setting, it could directly interface with the parallel
simulation, rather than having to save the data and reload it for analysis.
The issue of saving and reloading data is an important performance
bottleneck in distributed parallel computing, and has previously been addressed
in related work in the context of GPU parallelization~\cite{bernaschi17}.

Computing Voronoi cells in a distributed parallel environment is considerably
more challenging than the multithreaded case and requires different algorithms.
It is necessary to communicate to neighboring processors to obtain some of
their particle positions. For dense particle arrangements, it should usually be
sufficient for each processor to obtain particles in a region of fixed width
beyond its domain, and then use those to apply plane cuts during the Voronoi
cell construction. This is the approach taken by the \co{voronoi} command in
LAMMPS~\cite{lammps_website}. However, in some situations, such as when a
Voronoi cell extends out by a long way in one direction
(Fig.~\ref{fig:method_vorocell_costs}(b)), the fixed-width region will be
insufficient to completely determine the Voronoi cells. Because of this the
parallel implementation would also have to allow for the possibility of
non-local communication, where a processor could potentially talk to any other
to obtain its particle information. These are interesting parallel computation
challenges that can be explored in future work.

\section*{Acknowledgements}
This research was supported by a grant from the United States--Israel Binational
Science Foundation (BSF), Jerusalem, Israel through grant number 2018/170.
C.~H.~Rycroft was partially supported by the Applied Mathematics
Program of the U.S. DOE Office of Advanced Scientific Computing Research under
contract number DE-AC02-05CH11231. Additional support from the Data Science
Institute in Bar-Ilan University is also gratefully acknowledged.

\begin{appendices}
\counterwithin{figure}{section}
\section{Turbo Boost and parallel efficiency}
\label{appendix:turbo_boost}
Modern Intel CPUs make use of the Turbo Boost technology, which boosts the CPU
clock speed from its base value, depending on many factors (\textit{e.g.}\@ CPU
temperature and available power). Most significantly, Turbo Boost depends on
the number of cores in use. Our test system has dual Intel Xeon
E5-2650L v4 processors with a base clock speed of 1.7\,GHz.
Table~\ref{tbl:sec3_turbo_freq} shows the boosted clock speeds in terms of the
number of cores in use, ranging from 2.5\,GHz for a single core to 2.0\,GHz for
six or more cores.

\begin{table}
  \begin{center}
    \small
    \begin{tabular}{|c|c|c|}
      \hline
      Cores in use&Normal frequency each core (GHz)&Turbo Boost frequency each core (GHz)\\ \hline
      1--2& \multirow{5}{*}{1.7} &2.5\\ \cline{1-1} \cline{3-3}
      3& &2.3\\ \cline{1-1} \cline{3-3}
      4& &2.2\\ \cline{1-1} \cline{3-3}
      5& &2.1\\ \cline{1-1} \cline{3-3}
      6--14& &2.0\\
      \hline
    \end{tabular}
  \end{center}\vspace{-0.8em}
  \caption{Normal clock frequency and Turbo Boost frequency for Intel Xeon
  E5-2650L: v4 microprocessor used in the performance
  tests.}~\label{tbl:sec3_turbo_freq}
\end{table}

\begin{figure}
  \centering
  \includegraphics[width=0.8\textwidth]{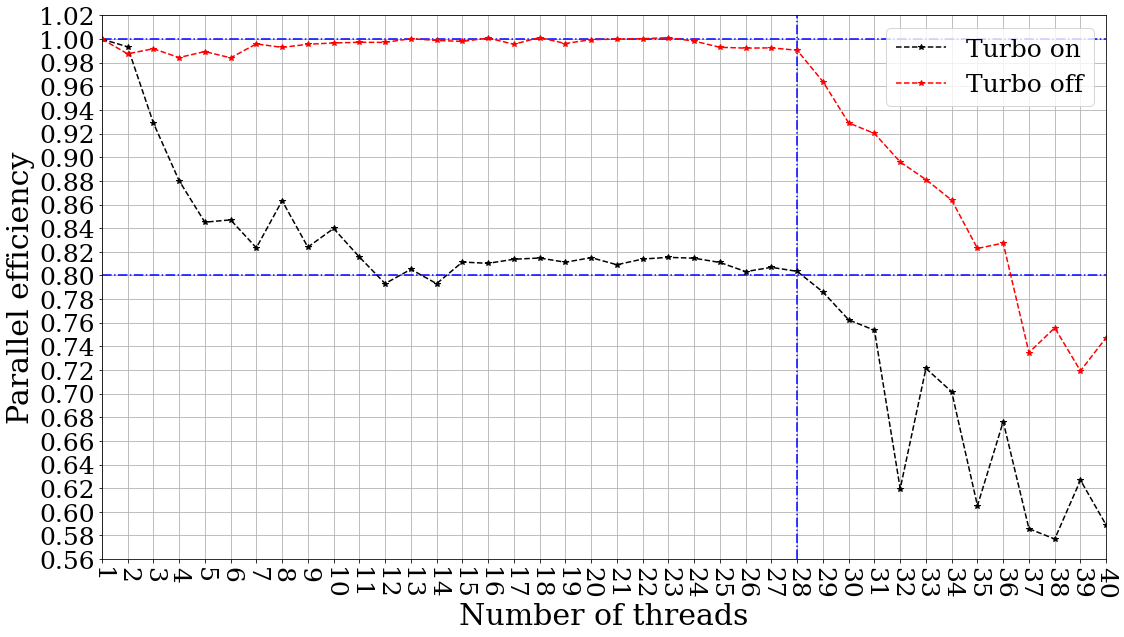}
  \caption{Comparison of parallel efficiencies against number of threads, with
  Turbo Boost on, and with Turbo Boost off. The data obtained is from the
  homogeneous particle distribution case in 3D, using its optimal strategy
  \co{schedule(guided)}. The data with Turbo Boost off is obtained from four
  runs of the same test. We take the minimum computation time of the four runs
  against number of threads, and calculate the parallel efficiency using the
  minimum time. This avoids small timing fluctuations and gives a more
  representative picture in parallel performance.~\label{fig:sec3_turbo}}
\end{figure}

\begin{figure}
  \centering
  \includegraphics[width=1\textwidth]{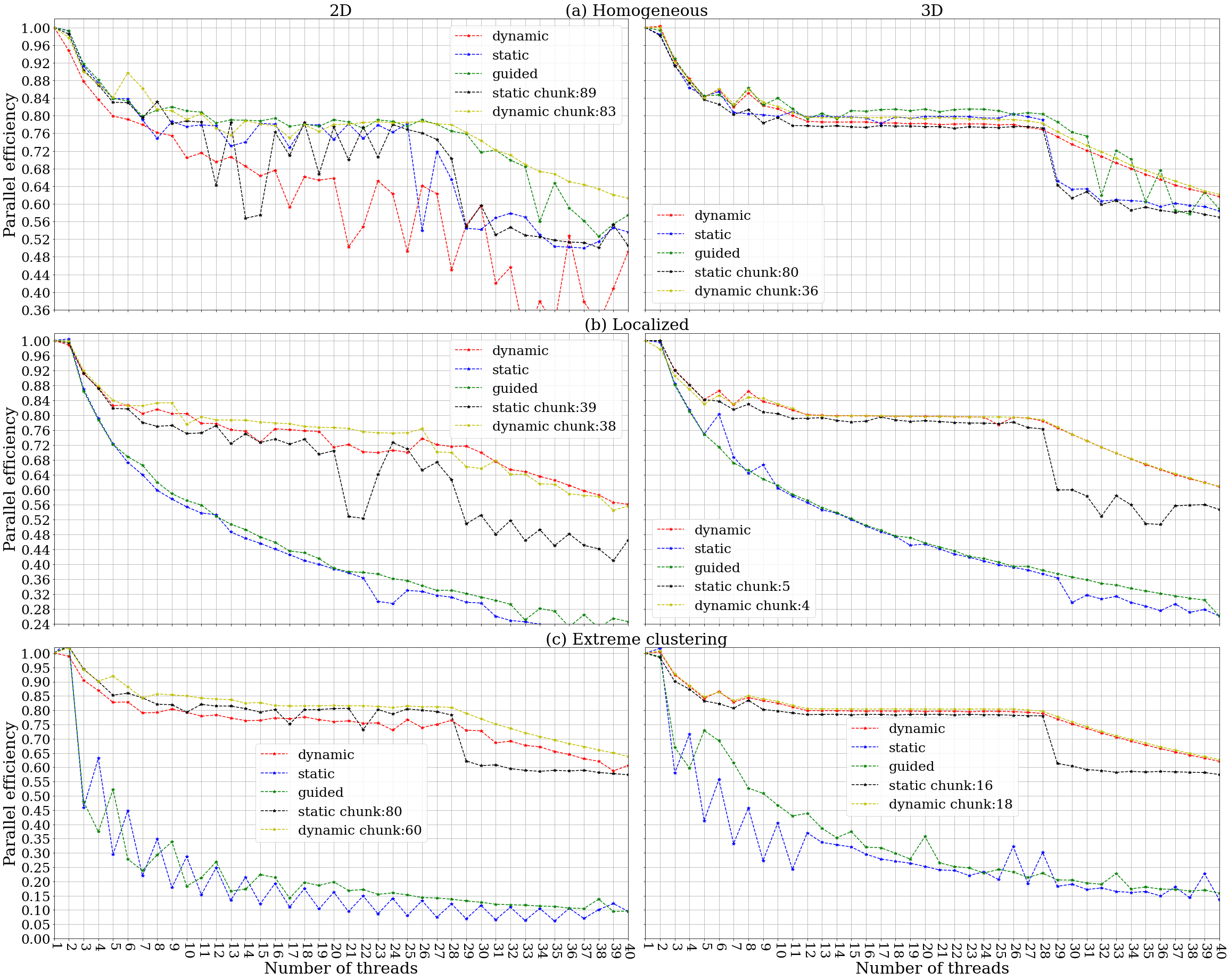}
  \caption{Unadjusted version of the parallel efficiency shown in
  Fig.~\ref{fig:sec3_eff_threads}, for the three particle arrangement cases in
  both 2D and 3D. (a) Homogeneous and random particle distribution. (b)
  Localized and random particle distribution. (c) Extreme clustering particle
  distribution.~\label{fig:unadj_eff_threads}}
\end{figure}

Turbo Boost has a noticeable effect on the measurement of parallel efficiency.
Figure~\ref{fig:sec3_turbo} shows the parallel efficiency for the random
homogeneous test case with $10^8$ particles with the \co{schedule(guided)}
strategy, evaluated using Eq.~\eqref{eq:orig_eff}. Once the number of threads
reaches 12, the parallel efficiency plateaus at $\approx 80\%$. This is
precisely in agreement with the ratio of clock speeds from 2.5\,GHz to
2.0\,GHz. This suggests that the reduction in efficiency is almost entirely
explained by Turbo Boost, rather than effects of the \vpp{} parallelization. To
verify this, we switched off Turbo Boost so that the clock speed remains
at 1.7\,GHz regardless of the number of cores in use.
Figure~\ref{fig:sec3_turbo} shows the parallel efficiency for this case. Even
though the simulation is slower overall, the parallel efficiencies remain close
to 100\%, confirming that Turbo Boost is responsible for the loss in parallel
efficiency. Across many of the timing results, Turbo Boost has a clear
signature in the unadjusted parallel efficiency, with plateaus at 80\% visible
(Fig.~\ref{fig:unadj_eff_threads}).

For the timing tests, we aim to use a measure of parallel efficiency that is
not closely tied to the computer hardware. One approach is to switch off Turbo
Boost completely, but this results in an overall loss of performance, and is
not reflective of typical computer usage. We therefore decided to adjust
the efficiency calculation to factor out the Turbo Boost as given in
Eq.~\eqref{eq:adj_eff}, scaling the usual measure by a factor of $A(p)$ for $p$
threads. Our first approach to construct the adjustment was by using the
frequencies of Table \ref{tbl:sec3_turbo_freq}, so that
\begin{equation}
\label{eqn:adj_factor_a}
A(p)=\frac{\text{boost frequency for a single thread}}{\text{boost frequency for $p$ threads}}.
\end{equation}
However, this approach is not suitable for the less efficient strategies. For
example, the \co{schedule(guided)} strategy of the 2D localized case has low
efficiencies. During this computation, some threads become idle while waiting
for other threads to finish their computation, and therefore the Turbo Boost
frequency changes depend on the number of active cores. Since the adjustment
in Eq.~\eqref{eqn:adj_factor_a} assumes that all threads remain active, it can
overestimate the parallel efficiency.

We therefore used a more general and accurate way to calculate the adjusted
parallel efficiency by replacing the boost frequencies in
Eq.~\eqref{eqn:adj_factor_a} with average clock speeds of the computation.
Average clock speed represents the true computational speed, and it is computed
by dividing the total unhalted core cycles by the total task-clock time.
Unhalted core cycles is a hardware performance counter, and can be measured for
each thread using \co{perf}, a Linux performance analyzing tool~\cite{perf}.
The number of unhalted core cycles for a thread captures the CPU cycles run on
the thread when the thread is active, which is an estimate of the work that the
thread contributes to the parallel computation of the program. Specifically, we
use \co{perf\_event\_open} with the \co{PERF\_COUNT\_HW\_CPU\_CYCLES}
configuration option for measurement~\cite{perf_event_open}. Task-clock time is
a software performance counter, and it can be measured for each thread using
the \co{PERF\_COUNT\_SW\_TASK\_CLOCK} configuration option with
\co{perf\_event\_open}~\cite{perf_event_open}. The task-clock time of a thread
reports a clock count specific to the task that is running on that thread.

We then measure unhalted core cycles $C_{\text{serial}}$ and the task-clock
time $T_\text{serial}^{\text{task}}$ that the computation uses for the serial code.
For the parallel code, we measure the unhalted cycles $C_i$ and the task-clock time
$T_{i}^{\text{task}}$ for each thread $i=1,\ldots,p$. Then we compute
\begin{align}
  \text{average clock speed for $1$ thread}&=\frac{C_{\text{serial}}}{T_\text{serial}^{\text{task}}}, \\
  \text{average clock speed for $p$ threads}&=\frac{\sum_{i=1}^p C_i}{\sum_{i=1}^pT_{i}^{\text{task}}},
\end{align}
which are used in Eq.~\eqref{eq:adj_factor} to compute the adjustment factor.

\end{appendices}

\bibliography{voro}

\end{document}